\documentclass[12pt]{article}

\def\slash#1{\setbox0=\hbox{$#1$}#1\hskip-\wd0\hbox to\wd0{\hss\sl/\/\hss}}
\usepackage{epsf, cite}
\usepackage{epsfig}                  
\usepackage[all]{xy}                 
\usepackage{amsfonts}
\usepackage{latexsym}
\usepackage{amsmath}
\usepackage{amssymb}
\usepackage{cite}

\usepackage{epsf, cite}

\makeatletter
\renewcommand\section{\@startsection {section}{1}{\z@}%
                                   {-3.5ex \@plus -1ex \@minus -.2ex}
                                   {2.3ex \@plus.2ex}%
                                   {\normalfont\large\bfseries}}
\renewcommand\subsection{\@startsection{subsection}{2}{\z@}%
                                     {-3.25ex\@plus -1ex \@minus -.2ex}%
                                     {1.5ex \@plus .2ex}%
                                     {\normalfont\bfseries}}
\makeatother

\def\lbldef#1#2{\expandafter\gdef\csname #1\endcsname {#2}}

\def\href#1#2{#2}

\def\beq{\begin{equation}}
\def\eeq{\end{equation}}
\def    \bea    {\begin{eqnarray}}
\def    \eea    {\end{eqnarray}}

\begin{document}
\pagestyle{plain}
\begin{titlepage}

\begin{center}

\hfill{QMUL-PH-2007-16} \\

\vskip 1cm

{{\Large \bf M-Theory Brane Deformations}} \\

\vskip 1.25cm {David S. Berman\footnote{email: D.S.Berman@qmul.ac.uk}
  and Laura C. Tadrowski\footnote{email: L.C.Tadrowski@qmul.ac.uk} }
\\
{\vskip 0.2cm
Queen Mary College, University of London,\\
Department of Physics,\\
Mile End Road,\\
London, E1 4NS, England\\
}

\end{center}
\vskip 1 cm

\begin{abstract}
\baselineskip=18pt\

Using the techniques developed by Lunin and Maldacena we calaculate
the supergravity solutions of membranes and fivebranes in the presence
of a background C field. All the distinct possible C-field configurations are
explored. Decoupling limits for these branes are then described that
preserve the deformation leading to families of M-theory brane
deformation duals. The decoupled geometry is then explored using probe
brane techniques and brane thermodynamics.
\end{abstract}

\end{titlepage}

\pagestyle{plain}

\baselineskip=19pt

\section{Introduction}
Over the past few years there has been a renewed interest in the 
various deformations
of the field theories that arise on brane world volumes. For the D3
brane, the undeformed world volume theory is $\mathcal{N}=4$ Yang-Mills. Its
deformations include: noncommutative Yang-Mills where the Neveu-Shwarz
two form is present along the brane world volume \cite{sw,mr,hi,aosj,cai}; the so called
{\it{dipole}} theory where the Neveu-Schwarz two form is present with
one leg along the brane world volume and one off it
\cite{bergman,ganor,iran1,iran2,karl}; 
and finally the $\mathcal{N}=1$ $\beta$-deformed
Yang-Mills theory where the Neveu Schwarz two form is present entirely off the
brane world volume \cite{lm}. Often these deformed theories have
interesting properties such as the S-duality properties of the
marginally deformed ${\mathcal{N}}=4$ 
theory \cite{dorey}. Other recent developments have been with so
called puff field theories \cite{ganor2}.

Many of the generalisations to M-theory have already been explored
though ofcourse even the undeformed case is somewhat of a mystery for
more than one conincident brane. The cases that have been examined
already are where the three form C-field is present with all legs
along the five-brane world volume \cite{berman1,troels} and the
generlisation of the dipole theory the so called discpole theory,
\cite{iran1,ganor3}. In this paper we systematically go through all the
distinct deformations for both the membrane and the five-brane with
the background C-field in all the
possible configurations with legs on and off the brane world
volume. (This will, of course, recover some of the already known
solutions). Essentially this will be a systematic application of the
method of Lunin and Maldacena  \cite{lm} to give all C-field deformations of
M-theory brane geometries.

With these geometries at hand we desribe various decouping limits
(where the deformation is preserved in the limit). We then explore
these decoupled geometries, which will be the supergravity duals of a deformed
M-theory brane world volume theory. Through use of probe brane techniques and an analysis
of the thermodynamic properties we will attempt to learn about these dual world volume theories.

Part of the motivation will also be to explore the solution generating
method and decoupling proceedures themselves. In particular, we 
will see how the solution generating technique and thus the
deformations of the dual theory and the decoupling limit commute. We
will also see how different embedding choices of the the solution
generating proceedure become relevant. Finally, we will also see
through thermodynamic calculations that some of the deformations will
leave the entropy invariant. This is a surprise given that in the dual
theory the deformation will alter the interactions.

Overall the calculations presented here are a laboratory to study how
decoupling limits and deformations of branes in M-theory work along
with the different aspects of hidden symmetries in string and M-theory.

There has also been the study of the dipole deformations of Yang-Mills
theory \cite{gursoy}
motivated by the idea that the deformation may provide an additional scale through
which additional decouplings may take place. The M-theory analogue of
this would also be interesting to explore expecially in the context of
$G_2$ compactification \cite{aw}.

\section{Deformation method}
This is a brief review of a technique, which can be used to construct solutions 
to eleven dimensional supergravity
with the C-field switched on. Essentially, it is the method, 
described by Lunin and Maldacena \cite{lm}. The origins of the method date back to the 
use of T-duality (perhaps combined with Lorentz transformations) as a
solution generating symmetry \cite{ortinetal}.
The starting point is an eleven dimensional background in which there
are at least three $U(1)$ isometries. These isometries will form a three
torus. The metric and the potential may then be
placed in the following adapted form:
\begin{eqnarray}
\mathrm{ds}^2&=&\Delta^{1/3}M_{ab}\mathcal{D}\varphi^a\mathcal{D}\varphi^b+\Delta^{-1/6}g_{\mu\nu}\mathrm{d}x^{\mu}\mathrm{d}x^{\nu}\\
C^{(3)}&=&\frac{1}{2}(C_{a\mu\nu}\mathcal{D}\varphi^a\wedge\mathrm{d}x^{\mu}\wedge{d}x^{\nu}+C_{ab\nu}\mathcal{D}\varphi^a\wedge\mathcal{D}\varphi^b\wedge\mathrm{d}x^{\nu})\nonumber\\
&+&\frac{1}{6}(C_{\mu\nu\lambda}\mathrm{d}x^{\mu}\wedge\mathrm{d}x^{\nu}\wedge\mathrm{d}x^{\lambda} + C_{abc}\mathcal{D}\varphi^a\wedge\mathcal{D}\varphi^b\wedge\mathcal{D}\varphi^c)\\
{\mathcal{D}}\varphi^a &{\equiv}&\mathrm{d}\varphi^a + {\mathcal{A}}_{\mu}^a\mathrm{d}x^{\mu}
\end{eqnarray}
where $a,b,c$ label the directions of the $T^3$ and
$\mu,\nu,\lambda\cdots$ are for the remaining coordinates. The
determinant of $M$ is one and so $\sqrt{\Delta}$ is the volume of the three-torus.
Performing an $S^1$ reduction along one of 
the $U(1)$'s (for illustrative purposes we here choose $\varphi_3$) produces a type IIA solution.
\begin{equation}
M_{ab}\mathcal{D}\varphi^a\mathcal{D}\varphi^b=e^{-2\phi/3}h_{mn}\mathcal{D}\varphi^m\mathcal{D}\varphi^n + e^{4\phi/3}(\mathcal{D}\varphi^3+N_m\mathcal{D}\varphi^m)^2
\end{equation} 
where $m$ and $n$ run over the remaining $U(1)$'s of the $T^3$ and
$h$, like $M$, has unit determinant. Then a T-Duality is performed on another of the $T^3$ cycles ($\varphi_1$ in what follows) and a toroidally compactified 
IIB solution is produced with 
$\{\varphi^1,\varphi^2\}$ as the coordinates of the two torus. The 
IIB solution is writen in terms of the original eleven dimensional
fields as follows, with IIB fields on the left and eleven dimensional
fields on the right:
\begin{eqnarray}
\mathrm{ds}^2&=&\frac{1}{h_{11}}\Big[\frac{1}{\sqrt{\Delta}}(D\varphi^1-CD\varphi^2)^2+\sqrt{\Delta}(D\varphi^2)^2\Big]+e^{2\phi/3}g_{\mu\nu}\mathrm{d}x^{\mu}\mathrm{d}x^{\nu}\\
B^{(2)}&=&\frac{h_{12}}{h_{11}}D\varphi^1\wedge D\varphi^2-C_{32\mu}D\varphi^2\wedge \mathrm{d}x^{\mu}+D\varphi^1\wedge\mathcal{A}^1-\frac{1}{2}C_{3\mu\nu}\mathrm{d}x^{\mu}\wedge\mathrm{d}x^{\nu}\nonumber\\
& \ & +C_{31\mu}\mathrm{d}x^{\mu}\wedge\mathcal{A}^1\\
e^{2\Phi}&=&\frac{e^{2\phi}}{h_{11}}, \ \ C^{(0)}=N_1\nonumber\\
C^{(2)}&=&-(N_{2}-\frac{h_{12}}{h_{11}}N_{1})D\varphi^1\wedge D\varphi^2-C_{12\mu}D\varphi^2\wedge\mathrm{d}x^{\mu}-D\varphi^1\wedge\mathcal{A}^3\nonumber\\
& &-\frac{1}{2}C_{1\mu\nu}\mathrm{d}x^{\mu}\wedge\mathrm{d}x^{\nu}+C_{31\mu}\mathrm{d}x^{\mu}\wedge\mathcal{A}^3\\
C^{(4)}&=&-\Big(\frac{1}{2}[C_{2\mu\nu}+2C_{32\mu}\mathcal{A}^3-\frac{h_{12}}{h_{11}}(C_{1\mu\nu}+2C_{31\mu}\mathcal{A}^3)]D\varphi^2\wedge\mathrm{d}x^{\mu}\wedge\mathrm{d}x^{\nu}\nonumber\\
& \ & +\frac{1}{6}(C_{\mu\nu\lambda}+3C_{3\mu\nu}\mathcal{A}^3)\mathrm{d}x^{\mu}\wedge\mathrm{d}x^{\nu}\wedge\mathrm{d}x^{\lambda}\Big)\wedge D\varphi^1 \nonumber\\
& \ & +d_{\mu_1\mu_2\mu_3\mu_4}\mathrm{d}x^{\mu_1}\wedge\mathrm{d}x^{\mu_2}\wedge\mathrm{d}x^{\mu_3}\wedge\mathrm{d}x^{\mu_4}\nonumber\\
& \ &
+\hat{d}_{\mu_1\mu_2\mu_3}\mathrm{d}x^{\mu_1}\wedge\mathrm{d}x^{\mu_2}\wedge\mathrm{d}x^{\mu_3}\wedge
D\varphi^2 \, .
\end{eqnarray} 
$D\varphi_1=\mathrm{d}\varphi_1-C_{31\mu}\mathrm{d}x^{\mu}, \ {\rm{and}} \ D\varphi_2=\mathrm{d}\varphi_2+\mathcal{A}_{\mu}^2\mathrm{d}x^{\mu}$. $C$ is the component $C_{123}$ on the $T^3$. The last two terms in the Ramond-Ramond four-form can be determined using
self-duality of the five-form field strength in ten dimensions. 
This reduction has an $SL(2,\mathcal{R})\times SL(2,\mathcal{R})$ 
symmetry which provide a means by which supergravity
solutions can be generated. The specific solution generating 
transformation described in \cite{lm} involves a T-Duality, a coordinate 
transformation and then another T-Duality. It is realised through the use of one of the 
$SL(2,\mathcal{R})$ symmetries to produce a rotation in the $\varphi^1-\varphi^2$ plane.
The particular element of $SL(2,\mathcal{R})$    
is chosen in such a way that the regularity of the solution is preserved so 
no new singular points are generated.  
In this way the symmetries of the eight dimensional theory are exploited
to generate new solutions of the eleven dimensional supergravity. 
The $SL(2,\mathcal{R})$ element used to produce the rotation is   
\[
\begin{pmatrix} \varphi^1 \\ \varphi^2\end{pmatrix}\longrightarrow
\begin{pmatrix} 1 & 0\\ \gamma & 1\end{pmatrix}\begin{pmatrix}
  \varphi^1 \\ \varphi^2\end{pmatrix} \, .
\]
Putting this rotation into effect in the toroidally compactified IIB theory
allows an alternative interpretation of this rotation as shifts in the fields of 11 
dimensional supergravity. Solutions obtained in this manner will be 
referred to as $\gamma$-deformed. It can be seen from the IIB metric that the effect of this
rotation in the $\varphi^1-\varphi^2$ plane is equivalently realised by making the following 
shifts in the eleven dimensional theory
\begin{eqnarray}
\label{deltaprime}\Delta_{\gamma} &=&\frac{\Delta}{[(1-\gamma C)^2+\gamma^2\Delta]^2}\\
\label{cprime}C_{\gamma} &=& \frac{C(1-\gamma
  C)-\gamma\Delta}{[(1-\gamma C)^2+\gamma^2\Delta]} \, .
\end{eqnarray}  
This analysis is considerably simplified for the branes in M-theory, which we will be discussing
because for every choice
of the $T^3$ used to generate the new solution, $C$
may be set to zero in the original solution\footnote{It is not that
  the C-field vanishes for these solutions just that the C field on
  the solution generating torus will vanish.}.
For some simple cases, (\ref{deltaprime}) and (\ref{cprime}) are the only changes required to prodce the new solution.
 There are, however, extra terms arising when, for example, 
the $T^3$ is completely in the space perpendicular to the
membrane world volume.

\section{The membrane}
We begin with the membrane. The undeformed solution is given by:
\begin
{eqnarray}
\mathrm{ds}^2&=&H^{-2/3}(-\mathrm{d}t^2+\mathrm{d}\rho^2+\rho^2\mathrm{d}\varphi_4^2)+H^{1/3}(\mathrm{d}r^2+r^2\mathrm{d}\Omega^2_7)\\
F_{0}^{(4)}&=&\rho\partial_rH^{-1}\mathrm{d}r\wedge\mathrm{d}t\wedge\mathrm{d}\rho\wedge\mathrm{d}\varphi_4
\, ,
\end{eqnarray}
with
\begin{equation}
H=1+ \frac{2^5 \pi^2 N l_p^6}{r^6} \, .
\end{equation}
We use the same coordinate system as Lunin and Maldacena to maintain
contact with their work\footnote{$c_{\theta}=\cos(\theta),
  s_{\theta}=\sin(\theta)$}. The volume element for the unit seven
sphere is given by 
\begin{eqnarray}
\mathrm{d}\Omega^2_7&=&\mathrm{d}\theta^2+s^2_\theta (\mathrm{d}\alpha^2+s^2_\alpha\mathrm{d}\beta^2)+c^2_{\theta}\mathrm{d}\phi_1^2\nonumber\\
&+&s^2_{\theta}(c^2_{\alpha}\mathrm{d}\phi_2^2+s^2_{\alpha}(c^2_{\beta}\mathrm{d}\phi_3^2+s^2_{\beta}\mathrm{d}\phi_4^2)))
\, .
\end{eqnarray}
One may then introduce the following coordinates \cite{lm}:
\begin{eqnarray}
\phi_1&=&\psi+\varphi_3\nonumber\\
\phi_2&=&\psi-\varphi_3-\varphi_2\nonumber\\
\phi_3&=&\psi+\varphi_2-\varphi_1\nonumber\\
\phi_4&=&\psi+\varphi_1 \, . \nonumber
\end{eqnarray} 
The background, adapted to begin the
deformation process described above, is:  
\begin{eqnarray}
\mathrm{ds}^2_0&=&H^{-2/3}\Big(-\mathrm{d}t^2+\mathrm{d}\rho^2+\rho^2\mathrm{d}\varphi_4^2\Big)\nonumber\\
&+&H^{1/3}\Big(\mathrm{d}r^2+r^2\big(\mathrm{d}\theta^2+s_{\theta}^2(\mathrm{d}\alpha^2+s_{\alpha}^2\mathrm{d}\beta^2)+\mathrm{d}\psi^2+s_{\theta}^2s_{\alpha}^2\mathrm{d}\varphi_1^2\nonumber\\
& &\ \ \ \ \ \ \ +2s_{\alpha}^2s_{\theta}^2(s_{\beta}^2-c_{\beta}^2)\mathrm{d}\psi\mathrm{d}\varphi_1+s_{\theta}^2(s_{\alpha}^2c_{\beta}^2+c_{\alpha}^2)\mathrm{d}\varphi_2^2\nonumber\\
& &\ \ \ \ \ \ \ +2s_{\theta}^2(s_{\alpha}^2c_{\beta}^2-c_{\alpha}^2)\mathrm{d}\psi\mathrm{d}\varphi_2-2c_{\beta}^2s_{\theta}^2s_{\alpha}^2\mathrm{d}\varphi_1\mathrm{d}\varphi_2\nonumber\\
& &\ \ \ \ \ \ \ +(c_{\theta}^2+s_{\theta}^2c_{\alpha}^2)\mathrm{d}\varphi_3^2+2(c_{\theta}^2-s_{\theta}^2c_{\alpha}^2)\mathrm{d}\psi\mathrm{d}\varphi_3+2s_{\theta}^2c_{\alpha}^2\mathrm{d}\varphi_2\mathrm{d}\varphi_3\big)\Big)\\
F^{(4)}_0 &=& \rho \, (\partial_{r}H^{-1}) \,
\mathrm{d}r\wedge\mathrm{d}t\wedge\mathrm{d}\rho\wedge\mathrm{d}\varphi_4
 \, .
\end{eqnarray}

There are two types of deformation to be considered for the membrane. 
The choice depends on whether we
wish to turn the C-field on completely off the brane
or with a single leg along the brane. 
In the former set-up the $T^3$ used for the solution generating transfromation 
will be transverse to the brane and in the latter, 
two of the directions of the torus
will be transverse and one will be longitudinal to the brane. 
There are four cases, which exhaust every possibile choice for both types of deformation. These four choices of $T^3$ are given by: 
\begin{displaymath}
\{\varphi^1,\varphi^2, \varphi^3\} \ \ \mathrm{or} \ \{\varphi^4,
\varphi^i, \varphi^j\}\ \ \mathrm{with} \ i,j=1,2,3 \, \, .
\end{displaymath}
Once the $T^3$ has been fixed, the particular choice of $S^1$ reduction direction and T-duality direction are
irrelevant in the eleven dimensional theory. However, the different embeddings of the specific $T^3$'s chosen will be shown to result in different solutions.

\subsection{Deformation for $T^3$ on $\{\varphi^1,\varphi^2, \varphi^3\}$}
The deformed M2 using $T^3$ given by $\{\varphi_1,\varphi_2, \varphi_3\}$ is  
\begin{eqnarray}
\mathrm{ds}^2&=&(1+\gamma^2\Delta_{123})^{1/3}\Bigg\{H^{-2/3}\Big(-\mathrm{dt}^2+\mathrm{d}\rho^2+\rho^2\mathrm{d}\varphi_{4}^2\Big)\nonumber\\
& &\ \ \ \ \ \ \ \ \ \ \ \ \ \ \ \ \ \ \ \ \ \ +H^{1/3}\Big(\mathrm{d}r^2+r^2\big(\mathrm{d}\theta^2+s_{\theta}^2(\mathrm{d}\alpha^2+s_{\alpha}^2\mathrm{d}\beta^2\big)\Big)\Bigg\}\nonumber\\
&+&\frac{H^{1/3}r^2}{(1+\gamma^2\Delta_{123})^{2/3}}\Bigg\{s_{\theta}^2s_{\alpha}^2\mathrm{d}\varphi_{1}^2+s_{\theta}^2(s_{\alpha}^2c_{\beta}^2+c_{\alpha}^2)\mathrm{d}\varphi_{2}^2\nonumber\\
& &\ \ \ \ \ \ \ \ \ \ \ \ \ \ \ \ \ \ \ \ \ \ \ +(c_{\theta}^2+s_{\theta}^2c_{\alpha}^2)\mathrm{d}\varphi_{3}^2\nonumber-2c_{\beta}^2s_{\theta}^2s_{\alpha}^2\mathrm{d}\varphi_{1}\mathrm{d}\varphi_{2}\nonumber\\
& &\ \ \ \ \ \ \ \ \ \ \ \ \ \ \ \ \ \ \ \ \ \ \ +2s_{\theta}^2c_{\alpha}^2\mathrm{d}\varphi_{2}\mathrm{d}\varphi_{3}+2s_{\alpha}^2s_{\theta}^2(s_{\beta}^2-c_{\beta}^2)\mathrm{d}\psi\mathrm{d}\varphi_{1}\nonumber\\
& & \ \ \ \ \ \ \ \ \ \ \ \ \ \ \ \ \ \ \ \ \ \ \ +2s_{\theta}^2(s_{\alpha}^2c_{\beta}^2-c_{\alpha}^2)\mathrm{d}\psi\mathrm{d}\varphi_{2}+2(c_{\theta}^2-s_{\theta}^2c_{\alpha}^2)\mathrm{d}\psi\mathrm{d}\varphi_{3}\Bigg\}\nonumber\\
&+&\frac{H^{1/3}r^2}{(1+\gamma^2\Delta_{123})^{2/3}}\Bigg\{\Big(1+\gamma^2\Delta_{123}f_1(\alpha,\beta,\theta)\Big)\mathrm{d}\psi^2\Bigg\}
\end{eqnarray}
where\footnote{\begin{eqnarray}
f_1(\alpha,\beta,\theta)&=&\frac{64\times256 \ \Delta'_{123} \ s_{2\alpha}^2s_{2\beta}^2s_{2\theta}^2}{s_{\alpha}^2s_{\theta}^4 \ f_2(\alpha,\beta,\theta)}\nonumber\\
f_2(\alpha,\beta,\theta)&=&66-2c_{4\beta}-3c_{4\beta-2\theta}-4c_{4(\beta-\theta)}+16c_{2\alpha}(7+c_{4\beta})c_{\theta}^2+70c_{2\theta}\nonumber\\
& &\ \ \ \ +8c_{4\theta}-4c_{4(\beta+\theta)}-3c_{2(2\beta+\theta)}+8c_{4\alpha}(7+8c_{2\theta})s_{2\beta}^2s_{\theta}^2
\end{eqnarray}} the volume of the three-torus is 

\begin{eqnarray}
\Delta_{123}&=&Hr^6s_{\alpha}^2s_{\theta}^4(c_{\beta}^2c_{\theta}^2s_{\alpha}^2s_{\beta}^2+c_{\alpha}^2(c_{\theta}^2+c_{\beta}^2s_{\alpha}^2s_{\beta}^2s_{\theta}^2))
\nonumber \\
&=&Hr^6\Delta'_{123}(\alpha,\beta,\theta)
\end{eqnarray}
There are two new contributions to the field strength. 
The first is the result of the three form potential components turned on along the $T^3$ during the 
deformation process. This can be seen by looking
at the effect of an $SL(2,\mathcal{R})$ rotation on the $T^2$ part 
of the IIB metric. Re-interpreting this transformation on the coordinates
as a shift in the fields of eleven dimensional supergravity gives
\begin{equation}
C_{\varphi_1\varphi_2\varphi_3}^{(\gamma)}=\frac{-\gamma\Delta}{1+\gamma^2\Delta}
\end{equation}
The second of the new contributions to the $\gamma$-deformed field strength 
originates with the non-zero $C_{t\rho\varphi_4}$. 
This enters the IIB theory through the Ramond-Ramond four form.
Specifically for this $T^3$ reduction, the RR field in the $T^2$ reduced IIB theory is
\begin{eqnarray}
C^{(4)}_0&=&-\frac{1}{6}C_{\mu\nu\lambda}\mathrm{d}x^{\mu}\wedge\mathrm{d}x^{\nu}\wedge\mathrm{d}x^{\lambda}\wedge D\varphi_1+d_{\mu_1\mu_2\mu_3\mu_4}\mathrm{d}x^{\mu_1}\wedge\mathrm{d}x^{\mu_2}\wedge\mathrm{d}x^{\mu_3}\wedge\mathrm{d}x^{\mu_4}\nonumber\\
& & +\hat{d}_{\mu_1\mu_2\mu_3}\mathrm{d}x^{\mu_1}\wedge\mathrm{d}x^{\mu_2}\wedge\mathrm{d}x^{\mu_3}\wedge D\varphi_2
\end{eqnarray}
Using the self-duality of the associated five-form field strength and
the isometries in the $\varphi_i$ directions implies that
\begin{equation}
d_{\mu_1\mu_2\mu_3\mu_4} = 0 \, .
\end{equation}
The $SL(2,\mathcal{R})$ action produces a new RR four-form
\begin{equation}
C^{(4)}_{\gamma}=C^{(4)}_{0}+\gamma \ \hat{d}_{\mu_1\mu_2\mu_3}\mathrm{d}x^{\mu_1}\wedge\mathrm{d}x^{\mu_2}\wedge\mathrm{d}x^{\mu_3}\wedge D\varphi^1
\end{equation}
Re-interpreting the $SL(2,\mathcal{R})$ transformation 
as a shift in the fields of 
eleven dimensional supergravity leads to the second new contribution to the eleven dimensional four-form
\begin{equation}
\frac{1}{6}C_{\mu\nu\lambda}\mathrm{d}x^{\mu}\wedge\mathrm{d}x^{\nu}\wedge\mathrm{d}x^{\lambda}\rightarrow\frac{1}{6}C_{\mu\nu\lambda}\mathrm{d}x^{\mu}\wedge\mathrm{d}x^{\nu}\wedge\mathrm{d}x^{\lambda}-{\gamma}
\
\hat{d}_{\mu_1\mu_2\mu_3}\mathrm{d}x^{\mu_1}\wedge\mathrm{d}x^{\mu_2}\wedge\mathrm{d}x^{\mu_3}
.
\end{equation}
The field strength for the deformed background is
\begin{equation}
F^{(4)}_{\gamma}=F^{(4)}_0-\partial_{\kappa} \ \Big[\frac{\gamma\Delta_{123}}{1+\gamma^2\Delta_{123}}\Big]\mathrm{d}\kappa\wedge\mathrm{d}\varphi_1\wedge\mathrm{d}\varphi_2\wedge\mathrm{d}\varphi_3+\gamma\sqrt{\Delta_{123}}\star_8F^{(4)}_0
\end{equation}
where $\kappa\in\{r,\alpha,\beta,\theta\}$.
The deformed solution, $ds^2_{(\gamma)}$ is related to the undeformed
solution, $ds^2_0$ by
\begin{equation}
\mathrm{ds}^2_{\gamma}=\frac{1}{(1+\gamma^2\Delta)^{2/3}} \ \Bigg\{\mathrm{ds}^2_0+\gamma^2\Delta\Big(\mathrm{d}x_{||}^2+H^{1/3}\sum_{i=1}^4\mathrm{d}\mu_i^2+r^2f_1\mathrm{d}\psi^2\Big)\Bigg\}
\end{equation}
where $\mathrm{d}x_{||}^2$ is the metric on the membrane world volume
\begin{equation}
\mathrm{d}x_{||}^2=H^{-2/3}\Big(-\mathrm{d}t^2+\mathrm{d}\rho^2+\rho^2\mathrm{d}\varphi_4^2\Big)
\end{equation}
with the $\mu_i$ coordinates, related to the chosen coordinate 
system by
\begin{equation}
\sum_{i=1}^4\mathrm{d}\mu_i^2=\mathrm{d}r^2+r^2\Big(\mathrm{d}\theta^2+s_{\theta}^2(\mathrm{d}\alpha^2+s_{\alpha}^2\mathrm{d}\beta^2)\Big)
\, .
\end{equation}
This shows, in an intuitive way, the effect of an $SL(2,\mathcal{R})$ rotation in this $T^2$ reduced IIB theory on the eleven dimensional geometry. For this specific deformation procedure, this is the only M-theory analogue of the $\beta$-deformation studied for the D3 Brane and it's world volume theory. In that case the deformation effect in the gauge theory was implemented
using a modified product for the Chiral Superfields. This is similar to the way in which a Non-Commutative theory can be obtained from a Commutative one after replacing the regular product with the Moyle product. However for the $\beta$-deformed D3/$\mathcal{N}=4$ system, the modified product produces an effect within the Chiral Superpotential alone. There appear multiplicative pure phase factors, which are determined by the charges of the gauge theory fields under the globally symmetric field theoretic realisation of the bulk spacetime torus. While the $\beta$-deformation preserved the conformal nature of the world volume theory, the Supersymmetry was broken down to $\mathcal{N}=1$. We investigate the effect of different $T^3$ embeddings on the Supersymmetry and thermodynamics in the M-theory set-ups.

For the membrane, a useful intrepretation is to view  the field theory
dual as the infrared (or equivalently strong coupling) limit of D=2+1
Yang-Mills theory. The $\gamma$ deformation will then be the analogue 
of the $\beta$ deformation
for the strongly coupled three dimensional theory. Deformations of 2+1
dimensional strongly coupled theories may have some condensed
matter applications, see \cite{sean} for recent uses of holography to
condensed matter systems.

\subsection{Membrane analogue of Dipole deformations}
For the second type of deformation of the membrane, the $T^3$ 
is chosen with one $U(1)$ on the brane world volume and two in the
transverse space. For all possible choices within this class of deformation the IIB
theory has Ramond-Ramond four form 
\begin{displaymath}
C^{(4)}=0 \, .
\end{displaymath}
The only new contribution to the eleven dimensional field strength comes from the 
potential turned on along the $T^3$. 
While there are other fields in the IIB
theory, which contain non-vanishing components of $C^{(3)}$, these can never generate new
eleven dimensional fields because they enter as $T^2$ zero-forms and two-forms only. 
The wedge product ensures that no new terms can be generated under the action of $SL(2,\mathcal{R})$ on the $T^2$ coordinates.  
Choosing to wrap the membrane on the $S^1$ producing a fundamental string in IIA appears to result
in the generation of novel new contributions.
However, the lack of the existence of a 
non-vanishing connection one-form for this choice of $U(1)$ results in  
this term always vanishing. Thus, all possible different IIB 
solutions originating from different choices of reduction and T-Duality angles are lifted
 to exactly the same solution in M-Theory. 
The deformed solutions are presented here in a form showing their relation to the undeformed membrane. The full solutions
can be found in Appendix A.
In general, a deformation of this type on $\{\varphi_4,\varphi_i,\varphi_j\}$ produces a solution of the form
\begin{eqnarray}
\mathrm{ds}^2_{\gamma}&=&\frac{1}{(1+\gamma^2\Delta_{4ij})^{2/3}}\Bigg\{\mathrm{ds}^2_0+\gamma^2\Delta_{4ij} \ \Bigg(H^{-2/3}(-\mathrm{d}t^2+\mathrm{d}\rho^2)\nonumber\\
& & \ \ \ \ \ \ +H^{1/3}\bigg(\sum_{i=1}^4\mathrm{d}\mu_i^2+r^2(\mathcal{F}_1 \ \mathrm{d}\psi^2+\mathcal{F}_2 \ \mathrm{d}\varphi_k^2+\mathcal{F}_3 \ \mathrm{d}\psi\mathrm{d}\varphi_k)\bigg)\Bigg)\Bigg\}\\
F^{(4)}_{\gamma}&=&F^{(4)}_0-\gamma \ \partial_{\kappa} \Big\{\frac{\Delta_{4ij}}{1+\gamma^2\Delta_{4ij}}\Big\} \ \mathrm{d}x^{\kappa}\wedge\mathrm{d}\varphi_4\wedge\mathrm{d}\varphi_i\wedge\mathrm{d}\varphi_j
\end{eqnarray}
where $\varphi_k$ labels the remaining $U(1)$ of the $T^3$,
$\kappa\in\{\rho,r,\alpha,\beta,\theta\}$ and
$\mathcal{F}_1,\mathcal{F}_2,\mathcal{F}_3$ are functions of
$\alpha,\beta,\theta $ only, depend on the specific $T^3$ chosen and
are presented in Appendix A. For each choice however, the function
$\Delta_{4ij}$ has the same $r$ dependence (allowing analysis of the
decoupling limits to be carried out in parallel) but different angular
dependence. This has consequences concerning the supersymmetry of the
deformed solution. The functions $\Delta_{4ij}$ are given by:
\begin{eqnarray}       
\Delta_{412}&=&\rho^2r^4s_{\theta}^4s_{\alpha}^2\Big[s_{\alpha}^2c_{\beta}^2+c_{\alpha}^2-c_{\beta}^4s_{\alpha}^2\Big]
\, ,\\
\Delta_{413}&=&\rho^2r^4s_{\theta}^2s_{\alpha}^2\Big[c_{\theta}^2+c_{\alpha}^2s_{\theta}^2\Big]
\, ,\\
\Delta_{423}&=&\rho^2r^4s_{\theta}^2\big[s_{\alpha}^2c_{\beta}^2(c_{\theta}^2+s_{\theta}^2c_{\alpha}^2)+c_{\alpha}^2c_{\theta}^2\Big]
\, .
\end{eqnarray}

\section{The five-brane}
The undeformed five-brane solution is given by:
\begin{eqnarray}
\mathrm{ds}^2&=&H^{-1/3}\Big(-\mathrm{d}t^2+\sum_{i=1}^{5}\mathrm{d}x_i^2\Big)+H^{2/3}\sum_{i=6}^{10}\mathrm{d}x_i^2\\
F^{(4)}&=&\frac{1}{4!}H^{2/3}\epsilon_{i_1i_2i_3i_4i_5} \
\partial^{i_5}H(x_6,\cdots,x_{10})\mathrm{d}x^{i_1}\wedge\mathrm{d}x^{i_2}\wedge\mathrm{d}x^{i_3}\wedge\mathrm{d}x^{i_4}
\, .
\end{eqnarray}
In what follows we will Wick rotate to time on the five-brane world
volume so that we may apply the usual solution generating
technique. We will then Wick rotate back at the end to give a good
solution. This is similar to the proceedure that was used to find the
supergravity duals to the noncommutative open string.
The isometry group on the M5 world volume after Wick rotation
is $SO(6)$. In the space orthogonal to the brane we have $SU(2)\times
SU(2)\approx SO(4)\subset SO(5)$ symmetry which may be used 
to identify two $U(1)$ isometries. 
\begin{equation}
x_6=r_1\mathrm{cos} \ \theta_1, \ \ x_7=r_1\mathrm{sin} \ \theta_1, \ \ x_8=r_2\mathrm{cos} \ \theta_2, \ \ x_9=r_2\mathrm{sin} \ \theta_2
\end{equation}
with
\begin{equation}
 x_{10}=\sqrt{r^2-r_1^2-r_2^2}  \, .
\end{equation}
The rescaling,
\begin{equation}
\tilde{r_1}=\frac{r_1}{r}, \ \ \tilde{r_2}=\frac{r_2}{r}
\end{equation}
is then performed to ensure that $r$ is the only ``radial'' type coordinate. The appearence of additional coordinate singularities at 
\begin{equation}
\tilde{r_1}^2+\tilde{r_2}^2=1
\end{equation}
is actually the single point
\begin{equation}
x_{10}=0
\end{equation}
and is the consequence of the lack of a globally well defined cover of
the sphere. This is a coordinate singularity and hence bengin.
The M5 solution after Wick rotation is then
\begin{eqnarray}    
\mathrm{ds}^2&=&H^{-1/3}\sum_{i=1}^{3}\Big(\mathrm{d}\mu_i^2+\mu_i^2\mathrm{d}\phi_i^2\Big)+H^{2/3}\Big\{\mathrm{d}r^2+r^2\Big(\nonumber\\
& &\ \ \ \ \ \ \ \ \ \ \ \ \ \ \ \tilde{r}_1^2\mathrm{d}\theta_1^2+\tilde{r}_2^2\mathrm{d}\theta_2^2+\frac{(1-\tilde{r}_2^2)}{(1-\tilde{r}_1^2-\tilde{r}_2^2)\textbf{}}\mathrm{d}\tilde{r}_1^2+\nonumber\\
& &\ \ \ \ \ \ \ \ \ \ \  \frac{(1-\tilde{r}_1^2)}{(1-\tilde{r}_1^2-\tilde{r}_2^2)}\mathrm{d}\tilde{r}_2^2+\frac{2\tilde{r}_1\tilde{r}_2}{(1-\tilde{r}_1^2-\tilde{r}_2^2)}\mathrm{d}\tilde{r}_1\mathrm{d}\tilde{r}_2\Big\}\\
F^{(4)}_0&=&\tilde{r}_1\tilde{r}_2r^4(\partial_{r}H)\mathrm{d}\tilde{r}_1\wedge\mathrm{d}\theta_1\wedge\mathrm{d}\tilde{r}_2\wedge\mathrm{d}\theta_2
\end{eqnarray}
with the harmonic function
\begin{equation}
H=\Big(1+\frac{N\pi l_p^3}{r^3}\Big) \, .
\end{equation}
We now use the Lunin/Maldacena 
parameterisation again for the $S^5$ and get
\begin{eqnarray}
\mathrm{ds}^2&=&H^{-1/3}\Bigg\{\mathrm{d}\rho^2+\rho^2\Big(\mathrm{d}\alpha^2+s_{\alpha}^2\mathrm{d}\beta^2+\mathrm{d}\psi^2+s_{\alpha}^2\mathrm{d}\varphi_1^2\nonumber\\
& &\ \ \ \ \ \ \ \ \ \ \ +(c_{\alpha}^2+s_{\alpha}^2c_{\beta}^2)\mathrm{d}\varphi_2^2+2s_{\alpha}^2(c_{\beta}^2-s_{\beta}^2)\mathrm{d}\psi\mathrm{d}\varphi_1\nonumber\\
& &\ \ \ \ \ \ \ \ \ \ \ +2(s_{\alpha}^2c_{\beta}^2-c_{\alpha}^2)\mathrm{d}\psi\mathrm{d}\varphi_2+2s_{\alpha}^2c_{\beta}^2\mathrm{d}\varphi_1\mathrm{d}\varphi_2\Big)\Bigg\}\nonumber\\
&+& \  H^{2/3}\Bigg\{\mathrm{d}r^2+r^2\Big(\tilde{r}_1^2\mathrm{d}\theta_1^2+\tilde{r}_2^2\mathrm{d}\theta_2^2+\frac{(1-\tilde{r}_2^2)}{(1-\tilde{r}_1^2-\tilde{r}_2^2)}\mathrm{d}\tilde{r}_1^2\nonumber\\
& &\ \ \ \ \ \ +\frac{(1-\tilde{r_1}^2)}{(1-\tilde{r}_1^2-\tilde{r}_2^2)}\mathrm{d}\tilde{r}_2^2+\frac{2\tilde{r}_1\tilde{r}_2}{(1-\tilde{r}_1^2-\tilde{r}_2^2)}\mathrm{d}\tilde{r}_1\mathrm{d}\tilde{r}_2\Big)\Bigg\}
\end{eqnarray}
At the level of the type IIB theory, choice of $S^1$-reduction and T-duality directions appear to produce different
solutions. The eleven dimensional lift of these solutions however
produce the same solution in M-theory. This is obviously a consequence
of eleven dimensional covariance that essentially produces a hidden
symmetry from the IIB point of view.

\subsection{M-theory analogue of Non-Commutative Deformation}
When the $T^3$ is chosen to lie completely on the Euclidean M5 world volume, 
the new contributions to the field strength originate with 
the same terms in the Ramond-Ramond four-form as for the membrane cases considered.
For this case, in a simple diagonal coordinate system, the deformed solution is 
\begin{eqnarray}
\mathrm{ds}^2_{\gamma}&=&(1+\gamma^2H^{-1}\mu_1^2\mu_2^2\mu_3^2)^{1/3}\Bigg\{H^{-1/3}\sum_{i=1}^3\mathrm{d}\mu_i^2+H^{2/3}\Big(\mathrm{d}r^2+r^2\mathrm{d}\Omega_4^2\Big)\Bigg\}\nonumber\\
&+&\frac{1}{(1+\gamma^2H^{-1}\mu_1^2\mu_2^2\mu_3^2)^{2/3}} \ \Bigg\{H^{-1/3}\sum_{i=1}^3\mu_i^2\mathrm{d}\phi_i^2\Bigg\}\\
& &\nonumber\\
F^{(4)}_{\gamma}&=&F^{(4)}_{0}+\gamma\sqrt{\Delta}\star_8F^{(4)}_{0}\nonumber\\
&-&\gamma\sum_{\kappa\in\{r,\mu_1,\mu_2,\mu_3\}}\partial_{\kappa}\frac{H^{-1}\mu_1^2\mu_2^2\mu_3^2}{(1+\gamma^2H^{-1}\mu_1^2\mu_2^2\mu_3^2)}\mathrm{d}x^{\kappa}\wedge\mathrm{d}\phi_1\wedge\mathrm{d}\phi_2\wedge\mathrm{d}\phi_3
\end{eqnarray}
where
\begin{equation}
\Delta_{\phi_1\phi_2\phi_3}=H^{-1}\mu_1^1\mu_2^2\mu_3^2  \, .
\end{equation}
The last term contributing to the new eleven dimensional field strength 
is from the potential component, which is turned on along the $T^3$ directions 
as a result of the deformation.
The other new contribution originates with the non-vanishing
$C_{\tilde{r_1}\theta_1\theta_2}$ and $C_{\tilde{r_2}\theta_1\theta_2}$. 
These live entirely in the 8 dimensional space 
perpendicular to the $T^3$ and appear in the $T^2$ reduced IIB theory
as non-vanishing $C_{\mu\nu\lambda}$. As for the membrane, this term is obtained
by transforming the eleven dimensional fields in such a way that they reproduce
the effect of an $SL(2,\mathcal{R})$ transformation in the IIB theory. 
\begin{equation}
F^{(4)}_{0}\rightarrow F^{(4)}_{0}+\gamma\sqrt{\Delta}\star_8F^{(4)}_{0}
\end{equation}

\subsection{M-theory analogue of Dipole Deformations on the five brane; type 1}
When two $U(1)$'s along the Brane are used (e.g $\phi_2,\phi_3$) along with one $U(1)$ 
from the perpendicular space (e.g $\theta_1$), the deformed solution is given by
\begin{eqnarray}
\mathrm{ds}^2_{\gamma}&=&\frac{1}{(1+\gamma^2\mu_2^2\mu_3^2r^2\tilde{r_1}^2)^{2/3}}\Bigg\{H^{-1/3}\sum_{i=2}^3\mu_i^2\mathrm{d}\phi_i^2+H^{2/3}r^2\tilde{r_1}^2\mathrm{d}\theta_1^2\Bigg\}\nonumber\\
&+&(1+\gamma^2\mu_2^2\mu_3^2r^2\tilde{r_1}^2)^{1/3}\Bigg\{H^{-1/3}\Big(\sum_{i=1}^3\mathrm{d}\mu_i^2+\mu_1^2\mathrm{d}\phi_1^2\Big)\nonumber\\
& &\ \ \ \ \ \ \ \ \ \ \ \ \ \ \ \ \ \ \ \ \  +H^{2/3}\Bigg(\mathrm{d}r^2+r^2\Big(\tilde{r_2}^2\mathrm{d}\theta_2^2+\frac{(1-\tilde{r_1}^2)}{(1-\tilde{r_1}^2-\tilde{r_2}^2)}\mathrm{d}\tilde{r_1}^2\nonumber\\
& &\ \ \ \ \ \ \ \ \ \ \ \ \ \ \ \ \ \ \ \ \  +\frac{(1-\tilde{r_1}^2)}{(1-\tilde{r_1}^2-\tilde{r_2}^2)}\mathrm{d}\tilde{r_2}^2+\frac{2\tilde{r_1}\tilde{r_2}}{(1-\tilde{r_1}^2-\tilde{r_2}^2)}\mathrm{d}\tilde{r_1}\mathrm{d}\tilde{r_2}\Big)\Bigg)\Bigg\}\\
F^{(4)}_{\gamma}&=&F^{(4)}_{0} - \gamma \sum_{\kappa}\partial_{\kappa} \ \frac{\mu_2^2\mu_3^2r^2 \tilde{r_1}^2}{(1+\gamma^2\mu_2^2\mu_3^2r^2\tilde{r_1}^2)} \mathrm{d}x^{\kappa}\wedge\mathrm{d}\phi_2\wedge\mathrm{d}\phi_3\wedge\mathrm{d}\theta_1
\end{eqnarray}
with $\kappa\in\{\mu_2,\mu_3, r, \tilde{r_1}\}$. From this solution, all others choices involving two $U(1)$'s ON and 
one $U(1)$ OFF the five brane world volume can be obtained via a
simple change of label amongst the $\phi_i, \ \mu_i \ i =1,2,3$ and
$\theta_m, \ \tilde{r_m}, \ m=1,2$. The lack of additional new terms
(common to all the alternative five brane 
deformations) is due to the existence of only a non-zero $C_{a\mu\nu}$
type term in the eleven dimensional solution. This appears only in
$T^2$ reduced IIB theory as zero-forms and two-forms on the $T^2$. No
new terms can be generated from these forms.

\subsection{M-theory analogue of Dipole Deformations on the five brane; type 2}
We now consider deforming the five brane using one world volume $U(1)$ and two from the transverse space. Unlike 
previous examples, it now matters whether the $S^1$ reduction and T-duality transformation are performed ON or OFF 
the brane world volume. We consider here three choices:

\begin{eqnarray}
1.&&S^1 \ \mathrm{reduce \ along \ brane \ w/v \ and \ T-dualise \ along \ transverse} \ U(1)\nonumber\\
2.&&S^1 \ \mathrm{reduce \ along \ transverse \ space \ and \ T-dualise \ along \ w/v} \ U(1)\nonumber\\
3.&&S^1 \ \mathrm{reduce \ and \ T-dualise  \ along \ transverse} \ U(1)\mathrm{'s}\nonumber
\end{eqnarray}
Each of these choices produces a toroidally compactified IIB solution with different Neveu-Schwarz Neveu-Schwarz two-forms and Ramond-Ramond two and four-forms. For the membrane no changes to the eleven dimensional three-form 
were obtained as the result of an $SL(2,\mathrm{R})$ transformation on the IIB solutions and it was concluded that 
the choice of T-duality and reduction directions were irrelevant from an eleven dimensional perspective. However, for the
five brane this is not the case. 

The third in the above list of choices is the easiest to deal with. T-dualising and $S^1$ reducing both along transverse 
directions results in the vanishing of all terms in the IIB theory, which are capable of generating new eleven 
dimensional field strength components through an $SL(2,R)$ rotation. We find for $T^3$ on $\{\varphi_1,\theta_1,\theta_2\}$

\begin{eqnarray}
B^{(2)}_{\gamma}&=&B^{(2)}_0=0\nonumber\\
C^{(2)}_{\gamma}&=&C^{(2)}_0=0\nonumber\\
C^{(4)}_{\gamma}&=&C^{(4)}_0=0\nonumber\\
\end{eqnarray}
and this is the same regardless of our choice of $S^1$-reduction and T-duality directions. The deformed solution obtained using choice 3 on
$\{\varphi_1,\theta_1,\theta_2\}$ is then

\begin{eqnarray}
\mathrm{ds}^2_{\gamma}&=&\frac{1}{(1+\gamma^2\Delta)^{2/3}}\Big(\mathrm{ds}^2_0+(\gamma^2\Delta) \ \Delta^{-1/6}g_{\mu\nu}\mathrm{d}x^{\mu}\mathrm{d}x^{\nu}\Big)\\
F^{(4)}_{\gamma}&=&F^{(4)}_0-\gamma\sum_{\kappa}\frac{\partial_{\kappa}(Hr^4\rho^2s_{\alpha}^2\tilde{r}_1^2\tilde{r}_2^2)}{(1+\gamma^2Hr^4\rho^2s_{\alpha}^2\tilde{r}_1^2\tilde{r}_2^2)^2}\mathrm{d}x^{\kappa}\wedge\mathrm{d}\phi_1\wedge\mathrm{d}\phi_2\wedge\mathrm{d}\theta_1
\end{eqnarray}
where $\kappa\in\{\rho,\alpha, r,\tilde{r_1},\tilde{r}_2\}$ and $\mu,\nu$ indices span the eight dimensional space. For the 
volume of the $T^3$ 
\begin{equation}
\Delta=\Delta_{\varphi_1\theta_1\theta_2}=Hr^4\tilde{r_1}^2\tilde{r_2}^2\rho^2s_{\alpha}^2
\end{equation}
and metric components 
\begin{eqnarray}
\Delta^{-1/6}g_{\mu\nu}&=&G_{\mu\nu} \ \ \mathrm{when} \ \mu,\nu \neq \{\varphi_2,\psi\}\nonumber\\
\Delta^{-1/6}g_{\varphi_2\varphi_2}&=&H^{-1/3}\rho^2\big(c_{\alpha}^2+s_{\alpha}^2c_{\beta}^2s_{\beta}^2\big)\nonumber\\
\Delta^{-1/6}g_{\psi\psi}&=&H^{-1/3}\rho^2\Big(1-s_{\alpha}^2\big(c_{\beta}^2-s_{\beta}^2\big)\Big)\nonumber\\
\Delta^{-1/6}g_{\varphi_2\psi}&=&2H^{-1/3}\rho^2\Big(2s_{\alpha}^2c_{\beta}^2s_{\beta}^2-c_{\alpha}^2\Big)\nonumber
\, .\\
\end{eqnarray}

For choices 1 and 2, however, there are more changes to be made. Whilst, in previous examples, the R-R four-form was responsible for newly generated terms, in these cases it produces no new effects. Instead, changes originate with
certain non-vanishing terms in the NS-NS and R-R two forms of the toroidally compctified IIB theory.

When the $S^1$ reduction is carried out along a $U(1)$ perpendicular to the brane world volume ($\theta_2$) and the solution is 
then T-dualised along a $U(1)$ on the world-volume ($\varphi_1$) the deformation in the IIB solution looks like
\begin{eqnarray}
B_{\gamma}^{(2)}&=&B_0^{(2)}+\gamma C_{\theta_1\theta_2\mu}\mathrm{D}\varphi_1\wedge\mathrm{d}x^{\mu}\nonumber\\
C_{\gamma}^{(2)}&=&C_0^{(2)}\nonumber\\
C_{\gamma}^{(4)}&=&C_0^{(4)} \, .
\end{eqnarray}
This can be re-interpreted as the following shift in the connection one-form for the T-dualised $U(1)$ direction
\begin{equation}
\mathcal{A}^{\varphi_1}_{0}\rightarrow\mathcal{A}^{\varphi_1}_{\gamma}=\mathcal{A}^{\varphi_1}_{0}+\gamma
C_{\theta_1\theta_2\mu}\mathrm{d}x^{\mu} \, .
\end{equation} 
The effect of choosing the other transverse $U(1)$ of the $T^3$ ($\theta_1$) as the $U(1)$ of the $S^1$ reduction
is the introduction of a negative sign due to the anti-symmetric nature of the three-form and we get
\begin{equation}
\mathcal{A}^{\varphi_1}_{0}\rightarrow\mathcal{A}^{\varphi_1}_{\gamma}=\mathcal{A}^{\varphi_1}_{0}-\gamma
C_{\theta_1\theta_2\mu}\mathrm{d}x^{\mu} \, .
\end{equation}
When the $S^1$ reduction is carried out along a $U(1)$ on the brane ($\varphi_1$) and 
the solution is then T-dualised along a $U(1)$ perpendicular to the world volume ($\theta_1$) the deformation in the IIB solution looks 
like 
\begin{eqnarray}
B_{\gamma}^{(2)}&=&B_0^{(2)}\nonumber\\
C_{\gamma}^{(2)}&=&C_0^{(2)}-\gamma C_{\theta_1\theta_2\mu}\mathrm{D}\theta_1\wedge\mathrm{d}x^{\mu}\nonumber\\
C_{\gamma}^{(4)}&=&C_0^{(4)}  \, .
\end{eqnarray}
This can be re-interpreted as the following shift in the connection one-form for the $U(1)$ of the $S^1$ reduction
\begin{equation}
\mathcal{A}^{\varphi_1}_{0}\rightarrow\mathcal{A}^{\varphi_1}_{\gamma}=\mathcal{A}^{\varphi_1}_{0}+\gamma
C_{\theta_1\theta_2\mu}\mathrm{d}x^{\mu} \, .
\end{equation} 
Again, we obtain a minus sign with the additional $\mathcal{O}(\gamma)$ terms when we choose to T-dualise along the other $U(1)$ ($\theta_2$)
of the $T^3$, which is perpendicular to the brane world volume as follows
\begin{equation}
\mathcal{A}^{\varphi_1}_{0}\rightarrow\mathcal{A}^{\varphi_1}_{\gamma}=\mathcal{A}^{\varphi_1}_{0}-\gamma C_{\theta_1\theta_2\mu}
\end{equation}
Thus, we find a symmetry in the following
pairs of solutions
\\
\\
Type A:
\begin{eqnarray}
&&S^1 \ \mathrm{Reduction \ on} \ \theta_2 \ \mathrm{with \ T-Duality \ along} \ \varphi_1\nonumber\\
&&S^1 \ \mathrm{Reduction \ on} \ \varphi_1 \ \mathrm{with \ T-Duality \ along} \ \theta_1\nonumber\\
\end{eqnarray}
Type B:
\begin{eqnarray}
&&S^1 \ \mathrm{Reduction \ on} \ \theta_1 \ \mathrm{with \ T-Duality \ along} \ \varphi_1\nonumber\\
&&S^1 \ \mathrm{Reduction \ on} \ \varphi_1 \ \mathrm{with \ T-Duality \ along} \ \theta_2\nonumber\\
\end{eqnarray}
with the only difference between the two pairs appearing as a relative minus sign in the $\mathcal{O}(\gamma)$ correction to the 
connection one form associated with the world volume $U(1)$.
For all choices involving a $T^3$ on $\{\varphi_1,\theta_1,\theta_2\}$ the deformed field strength is given 
by
\begin{equation}
F^{(4)}_{\gamma}=F^{(4)}_{0}-\gamma\sum_{\kappa} \ \frac{\partial_{\kappa}\Delta}{(1+\gamma^2\Delta)^2}\mathrm{d}x^{\kappa}\wedge\mathrm{d}\varphi_1\wedge\mathrm{d}\theta_1\wedge\mathrm{d}\theta_2
\end{equation}
with $\kappa\in\{\rho,\alpha,r,\tilde{r_1},\tilde{r_2}\}$ and
\begin{equation}
\Delta=\Delta_{\varphi_1\theta_1\theta_2}=H\rho^2s_{\alpha}^2r^4\tilde{r_1}^2\tilde{r_2}^2
\, .
\end{equation}
The metric is given by 
\begin{eqnarray}
\mathrm{ds}^2_{\gamma}&=&\frac{1}{(1+\gamma^2\Delta)^{2/3}}\Big\{\mathrm{ds}^2_{0}+\Delta^{-1/6}(\gamma^2\Delta)g_{\mu\nu}\mathrm{d}x^{\mu}\mathrm{d}x^{\nu}\nonumber\\
& &\ \ \ \ \ \ \ \ \ \ \ \ \ \ \ \ \ \ +\Delta^{1/3}M_{\varphi_1\varphi_1}(\delta^{\varphi_1})^2+2(-1)^P  \Delta^{1/3}M_{\varphi_1\varphi_1}\mathcal{D}\varphi^1\delta^{\varphi^1}\Big\}
\end{eqnarray}
where the last term is the only one for which there is a difference between the metrics produced by
the two types of deformation described here. 
The diference is given by
\begin{eqnarray}
P&=&1  \ \mathrm{for \ type \ A \ deformations}\nonumber\\ 
P&=&2  \ \mathrm{for \ type \ B \ deformations}\nonumber
\end{eqnarray}
Common to both types of deformation are
\begin{eqnarray}
\Delta^{-1/6}g_{\mu\nu}&=&G^{0}_{\mu\nu} \ \mathrm{for} \ \mu,\nu\notin\{\psi,\varphi_2\}\nonumber\\
\Delta^{-1/6}g_{\psi\psi}&=&H^{-1/3}\rho^2\Big(1-s_{\alpha}^2(c_{\beta}^2-s_{\beta}^2)^2\Big)\nonumber\\
\Delta^{-1/6}g_{\varphi_2\varphi_2}&=&H^{-1/3}\rho^2\Big(c_{\alpha}^2+s_{\alpha}^2c_{\beta}^2s_{\beta}^2\Big)\nonumber\\
\Delta^{-1/6}g_{\psi\varphi_2}&=&H^{-1/3}\rho^2\Big(s_{\alpha}^2c_{\beta}^2-c_{\alpha}^2-c_{\beta}^2(s_{\beta}^2-c_{\beta}^2)\Big)
\, .
\nonumber
\end{eqnarray}
The $(\delta^{\varphi_1})^2$ terms, which, along with the new off-diagonal components generated as a result of the shift in the connection one-form appear in terms of
\begin{eqnarray}
\delta^{\varphi_1}&=&\mathcal{A}^{\varphi_1}_{\gamma}-\mathcal{A}^{\varphi_1}_{0}\nonumber\\
&=&\mathcal{A}^{\varphi_1}_{\tilde{r_1}}\mathrm{d}\tilde{r_1}+\mathcal{A}^{\varphi_1}_{\tilde{r_2}}\mathrm{d}\tilde{r_2}\nonumber\\
&=&\gamma C_{\tilde{r_1}\theta_1\theta_2}\mathrm{d}\tilde{r_1}-\gamma C_{\theta_1\tilde{r_2}\theta_2}\mathrm{d}\tilde{r_2}\nonumber\\
&=&\frac{1}{4}\gamma r^4(\partial_r H) \
\Big(\tilde{r_1}\tilde{r_2}^2\mathrm{d}\tilde{r_1}-\tilde{r_1}^2\tilde{r_2}\mathrm{d}\tilde{r}_2
\Big) \, ,
\end{eqnarray}
with 
\begin{equation}
\mathcal{D}\varphi_1=\mathrm{d}\varphi_1+(c_{\beta}^2-s_{\beta}^2)\mathrm{d}\psi
+ c_{\beta}^2\mathrm{d}\varphi_2 \, .
\end{equation}
We could, of course, have chosen to use an \{ON,OFF,OFF\} \ $T^3$
involving $\varphi_2$ instead of $\varphi_1$. The effect of this
choice follows trivially from the solution in terms of
$\varphi_1$. The deformed field strength is given by direct exchange
of $\varphi_2$ for $\varphi_1$ in components. The volume of the $T^3$
changes only by an overall factor determined by the undeformed metric
as follows:
\begin{equation}
\Delta_{\varphi_2\theta_1\theta_2}=\frac{G^{0}_{\varphi_2\varphi_2}}{G^{0}_{\varphi_1\varphi_1}}\Delta_{\varphi_1\theta_1\theta_2}
\, .
\end{equation}
The changes to the metric are fully contained within the following interchanges
\begin{eqnarray}
\mathcal{D}\varphi^1&\rightarrow&\mathcal{D}\varphi^2\nonumber\\
\mathcal{A}_{\gamma}^{\varphi_1}&\rightarrow&\mathcal{A}_{\gamma}^{\varphi_2}\nonumber\\
\end{eqnarray}
whilst, as for the $\varphi_1$ choice 
\begin{equation}
\mathcal{A}_{\gamma}^{\varphi_2}=\mathcal{A}^{\varphi_2}_{0}+\delta^{\varphi_2}
\, .
\end{equation}
Obviously, as this is a different embedding
\begin{eqnarray}
\mathcal{A}_{0}^{\varphi_2}&=&\mathcal{A}_{\psi}^{\varphi_2}\mathrm{d}\psi+\mathcal{A}_{\varphi_1}^{\varphi_2}\mathrm{d}\varphi_1\nonumber\\
&=&\frac{s_{\alpha}^2c_{\beta}^2-c_{\alpha}^2}{s_{\alpha}^2c_{\beta}^2+c_{\alpha}^2}\mathrm{d}\psi+\mathcal{A}_{\varphi_2}^{\varphi_1}\mathrm{d}\varphi_1\nonumber\\
&\neq&\mathcal{A}_{0}^{\varphi_1}
\end{eqnarray}
but 
\begin{equation}
\delta^{\varphi_2}=\delta^{\varphi_1} \, .
\end{equation}
The result of these symmetries between the deformed solution is that the r-dependence of the metric and potential components are the same for all possible choices of deformation after we have specified two pieces of information. Firstly, the type of brane 
we are working with and secondly, the number of $U(1)$'s, which 
are parallel and transverse to the brane world volume. The analysis of the near horizon regions can therefore be carried out for many deformation cases simulataneously.

\section{Decoupling}

The Supergravity action tells us how to take a sensible decoupling
limit. As the only parameter in eleven 
dimensional supergravity, the planck length 
is used to determine the low energy limit, by taking $l_p \rightarrow
0$ relative to some fixed energy scale. We then
scale r appropriately so the supergravity action is finite in the low
energy limit.

The Bosonic part of the 
action is
\begin{eqnarray}
\mathcal{S}&=&\mathcal{S}_1+\mathcal{S}_2+\mathcal{S}_3\\
&=&\frac{1}{\kappa_{11}^2}\int\mathrm{d}^{11}x\,\Big\{\sqrt{-g}(\mathcal{R}-\frac{1}{12}\mid F\mid^2)+\frac{2}{(72)^2}\epsilon^{M_1\cdots M_{11}}F_{M_1\cdots M_4}F_{M_5\cdots M_8}C_{M_9M_{10}M_{11}}\Big\}\nonumber
\end{eqnarray}
where the eleven dimensional gravitational coupling is
\begin{equation}
\kappa_{11}^2\sim l_p^9  \, .
\end{equation}

\subsection{The membrane}

For the membrane 
the decoupling limit is taken by
\begin{equation}
l_p\rightarrow 0
\end{equation}
with
\begin{equation}
u^{1/2}\equiv \frac{r}{l_p^{3/2}}=\mathrm{fixed} \, .
\end{equation}
This is the near horizon limit defined by Maldacena. 
For the undeformed membrane an $AdS_4\times S^7$ near horizon geometry 
is produced. By comparing the near horizon of the deformed M2 background 
with the deformed $AdS_4\times S^7$ we can investigate the relationship between
of the deformation process and the near horizon limit.    
The three cases where the deformation is performed using a $T^3$ with
one cycle on the brane world volume and two cycles in the transverse space
have $\Delta$ factors with the same r-dependence 
They also have the same r-dependence in the non-vanishing $C^{(3)}$ 
components. Therefore, their near horizon limits can be analyzed simulataneously. However, the
case with all $U(1)$'s of the $T^3$ off the membrane world volume must be 
treated seperately. Importantly we shall also need to scale $\gamma$
so that the deformation is not washed out by the limit. 
 
The deformed membranes to be considered explicitly are those with $T^3$
given by $\{\varphi_1,\varphi_2,\varphi_3\}$ and 
$\{\varphi_4,\varphi_1,\varphi_3\}$. In both cases, there appear several different $r$-dependent combinations 
in the metric. For $\{\varphi_1,\varphi_2,\varphi_3\}$ these are
\begin{displaymath}
H^{-2/3}(1+\gamma^2\Delta_{123})^{1/3}, \ \ H^{1/3}r^2(1+\gamma^2\Delta_{123})^{1/3},\ \ \frac{H^{1/3}r^2}{(1+\gamma^2\Delta_{123})^{2/3}}
\end{displaymath}
where 
\begin{equation}
\Delta_{123}=Hr^6\Delta'_{123}(\alpha,\beta,\theta) \, ,
\end{equation}
with $\delta'$ independent of r. For the $\{\varphi_4,\varphi_1,\varphi_3\}$ deformation the
metric contains
\begin{displaymath}
\frac{H^{-2/3}}{(1+\gamma^2\Delta_{413})^{2/3}}, \ \ H^{2/3}(1+\gamma^2\Delta_{413})^{1/3},\ \ \frac{H^{1/3}r^2}{(1+\gamma^2\Delta_{413})^{2/3}},\ \ H^{1/3}r^2(1+\gamma^2\Delta_{413})^{1/3}
\end{displaymath}
with
\begin{equation}
\Delta_{413}=\rho^2 r^4 \Delta'_{413}(\alpha,\beta,\theta) \, .  \nonumber
\end{equation}
Finiteness of the Einstein-Hilbert term requires
\begin{equation}
\frac{\mathrm{ds}^2}{l_p^2}\equiv\mathrm{finite} \, .
\end{equation}
For all possible deformations we must scale for $\gamma$ in such a way
as to preserve the effect of the deformation in the near horizon
region. This implies we must keep
\begin{equation}
(1+\gamma^2\Delta)\equiv\mathrm{fixed}
\end{equation}
while
\begin{eqnarray}
H&\sim&\frac{(2^5\pi^2N)}{u^3l_p^3} \, .
\end{eqnarray}
For $T^3$ given by $\{\varphi_1,\varphi_2,\varphi_3\} $ we find 
\begin{equation}
\gamma^2Hr^6\equiv\mathrm{fixed}
\end{equation}
in the near horizon limit. Changing to the new $u$ variable and allowing
$l_p\rightarrow 0$
gives
\begin{equation}
\gamma^2l_p^6\equiv\mathrm{finite}
\end{equation}
 and so the deformation parameter is scaled as
\begin{equation}
\gamma\thicksim\tilde{\gamma}l_p^{-3}
\end{equation}
where $\tilde{\gamma}$ is a scale independent deformation parameter. 
Similarily, for the deformation involving 
$T^3$ on $\{\varphi_4,\varphi_1, \varphi_3\} $ the scale independent 
combination must be
\begin{equation}
\gamma^2r^4\equiv\mathrm{fixed} \, .
\end{equation} 
In the near horizon limit $\gamma$ is scaled like
\begin{equation}
\gamma\thicksim\tilde{\gamma}l_p^{-3} \, .
\end{equation}
This particular scaling is common to both types of deformation and   
it keeps the Einstein-Hilbert term of the eleven dimensional supergravity 
action finite in the near horizon region as required. 
We must also check that the terms invoving the C
field are also finite in this limit. This is not ensured since the
deformation switches on new values components of C and its associated
field strength. since the deformation paprameter must scale so will
the field strength and so we must check the action remains finite.

From the undeformed membrane
\begin{eqnarray}
\mathcal{S}_2&=&\frac{1}{\kappa_{11}^2}\int\,\mathrm{d}^{11}x\sqrt{-g}\mid F^{(4)}\mid^2\\
&\sim&\frac{1}{l_p^9}\int\,\mathrm{d}^{11}x\underbrace{\sqrt{-g}}_{\sim l_p^{11}}F_{ut\rho\varphi_4}\underbrace{g^{tt}g^{\rho\rho}g^{\varphi_4\varphi_4}g^{uu}}_{\sim (l_p^{-2})^4}F_{ut\rho\varphi_4}+\cdots
\end{eqnarray} 
Looking at the undeformed field strength in the near horizon limit we have
\begin{eqnarray}
F_{ut\rho\varphi_4}&=&\frac{\mathrm{d}r}{\mathrm{d}u}\rho \ \partial_{r}H^{-1}\nonumber\\
&=&\frac{l_p^{3/2}}{2\sqrt{u}}\,\rho\,\frac{1}{H^2}\frac{6(2^5\pi^2N)l_p^6}{r^7}\nonumber\\
&{\sim\atop{l_p\rightarrow 0}}&l_p^3 \ \Big[\frac{3\rho u^2}{(2^5\pi^2N)}\Big]
\end{eqnarray}
where the factor in brackets is fixed in the near horizon limit. 
This $l_p$ dependence is just the one required to ensure the action 
is finite in the decoupling limit provided.
For deformation on $T^3$ given by $\{\varphi_1,\varphi_2,\varphi_3\}$
non-zero 
$F_{\kappa\varphi_1\varphi_2\varphi_3}$
is generated. These components have two
possible $r$-dependencies. When $\kappa=r$ we get
\begin{equation}
F_{r\varphi_1\varphi_2\varphi_3}=-\frac{\gamma \
  (6r^5)\Delta'_{123}(\alpha,\beta,\theta)}{(1+\gamma^2Hr^6\Delta'_{123}(\alpha,\beta,\theta))^2} \, .
\end{equation} 
In the near horizon limit, the denominator is fixed and we see
\begin{eqnarray}
F_{u\varphi_1\varphi_2\varphi_3}&{\sim\atop{l_p\rightarrow 0}}& l_p^6
\
\Big[\frac{-3\tilde{\gamma}\Delta'_{123}u^2}{(1+\tilde{\gamma}^2(2^5\pi^2N)\Delta'_{123})^2}\Big]
\, .
\end{eqnarray}
Since only contributions to the field strength, which are finite in units of $l_p^3$ affect the 
near horizon region the effects of $F_{u\varphi_1\varphi_2\varphi_3}$ vanishes there.  
Alternatively, when $\kappa\in\{\alpha,\beta,\theta\}$
\begin{equation}
F_{\kappa\varphi_1\varphi_2\varphi_3}=\frac{-\gamma Hr^6 \partial_{\kappa}\Delta'_{123}}{(1+\gamma^2\Delta_{123})^2}
\end{equation} 
which scales like
\begin{eqnarray}
\gamma Hr^6 &{\sim\atop{l_p\rightarrow 0}} \ \ \ l_p^3 \ \Big[\tilde{\gamma}(2^5\pi^2N)\Big]
\end{eqnarray}
as required for it to remain present in the near horizon limit.
Another contribution to the action comes from
\begin{equation}
\mathcal{S}_2=\cdots+\frac{1}{\kappa_{11}^2}\int\,\mathrm{d}^{11}x\,\sqrt{-g}F_{\alpha\beta\theta\psi}g^{\alpha\alpha}g^{\beta\beta}g^{\theta\theta}g^{\psi\psi}\, F_{\alpha\beta\theta\psi}
\end{equation}  
where $F_{\alpha\beta\theta\psi}$ is one of the components of the field
generated in the Ramond-Ramond four-form of IIB when we performed an 
$SL(2,\mathcal{R})$ transformation. It was found to be 
\begin{eqnarray}
F_{\alpha\beta\theta\psi}\mathrm{d}\alpha\wedge\mathrm{d}\beta\wedge\mathrm{d}\theta\wedge\mathrm{d}\psi=\gamma\sqrt{\Delta_{123}} \ \star_8F_{0}^{(4)}
\end{eqnarray}  
where $F_0^{(4)}$ is the field strength for the undeformed M2 and the hodge
star is in eight dimensions. The symmetry in the appearence of $F_{ut\rho\varphi_4}$ and
$F_{\alpha\beta\theta\psi}$ indicates that to keep the $F_{\alpha\beta\theta\psi}$ we require the same near horizon
scalings

\begin{equation}
 F_{0}^{(4)}\sim\gamma\sqrt{\Delta_{123}} \ \star_8 F_{0}^{(4)}\sim l_p^3
\end{equation}
The eight dimensional hodge star involves the square root of the 
determinant of the eight dimensional metric along with the product 
of four inverse metrics. As a consequence of the finiteness of the 
Hilbert term in the action, the latter contributes a total of $(l_p)^{-8}$ 
while the former contributes $(l_p)^8$. At the same time
\begin{eqnarray}
\sqrt{\Delta_{123}}&{\sim\atop{l_p\rightarrow 0}}&  \ l_p^3 \ \Big[(2^5\pi^2N)^{1/2} \sqrt{\Delta'_{123}}\Big]
\end{eqnarray}  
This combines with the $\gamma$ factor to produce a scale independent
quantity giving in the near horizon
\begin{equation}
\underbrace{\gamma\sqrt{\Delta_{123}}}_{l_p^0} \ \underbrace{\sqrt{\mid g_{8}\mid} \ g^{uu}g^{tt}g^{\rho\rho}g^{\varphi_4\varphi_4}}_{l_p^0} \sim l_p^0
\end{equation}
and thus
\begin{equation}
\mathcal{S}_2=\frac{1}{2\kappa_{11}^2}\int\,\mathrm{d}^{11}x\,\sqrt{-g}\,\mid F^{(4)}\mid 
\end{equation}
is finite in the near horizon limit for
\begin{eqnarray}
l_p&\rightarrow&0\nonumber\\
u&=&\mathrm{fixed}\nonumber\\
\gamma&\sim&\tilde{\gamma}\, l_p^{-3} \, .
\end{eqnarray}
Having defined a consistent near horizon limit with finite action for which the deformation is preserved we now present the solutions. The $\gamma$-deformed membrane (with the $T^3$ in the transverse space)
in the near horizon is 
\begin{eqnarray}
\mathrm{ds}_{\gamma}^2&=&l_p^2\Bigg[(1+\tilde{\gamma}^2(2^5\pi^2N)\Delta'_{123})^{1/3}\Big\{\frac{u^2}{(2^5\pi^2N)^{2/3}}(-\mathrm{d}t^2+\mathrm{d}\rho^2+\rho^2\mathrm{d}\varphi_4^2)\nonumber\\
& &\ \ \ \ \ \ \ \ \ \ \ \ \ \ \ \ \ \ \ \ \ \ \ \ +(2^5\pi^2N)^{1/3}\Big(\frac{\mathrm{d}u^2}{4u^2}+\mathrm{d}\theta^2+s_{\theta}^2(\mathrm{d}\alpha^2+s_{\alpha}^2\mathrm{d}\beta^2)\Big)\Big\}\nonumber\\
& & \ \ \  +\frac{(2^5\pi^2N)^{1/3}}{(1+\tilde{\gamma}^2(2^5\pi^2N)\Delta'_{123})^{2/3}}\Big\{s_{\theta}^2s_{\alpha}^2\mathrm{d}\varphi_1^2+s_{\theta}^2(s_{\alpha}^2c_{\beta}^2+c_{\alpha}^2)\mathrm{d}\varphi_2^2\nonumber\\
& &\ \ \ \ \ \ \ \ \ \ \ \ \ \ \ \ \ \ \ \ \ \ \ \ +(c_{\theta}^2+s_{\theta}^2c_{\alpha}^2)\mathrm{d}\varphi_3^2-2c_{\beta}^2s_{\theta}^2s_{\alpha}^2\mathrm{d}\varphi_1\mathrm{d}\varphi_2\nonumber\\
& &\ \ \ \ \ \ \ \ \ \ \ \ \ \ \ \ \ \ \ \ \ \ \ \ +2s_{\theta}^2c_{\alpha}^2\mathrm{d}\varphi_2\mathrm{d}\varphi_3+2s_{\alpha}^2s_{\theta}^2(s_{\beta}^2-c_{\beta}^2)\mathrm{d}\varphi_1\mathrm{d}\psi\nonumber\\
& &\ \ \ \ \ \ \ \ \ \ \ \ \ \ \ \ \ \ \ \ \ \ \ \ +2s_{\theta}^2(s_{\alpha}^2c_{\beta}^2-c_{\alpha}^2)\mathrm{d}\varphi_2\mathrm{d}\psi+2(c_{\theta}^2-s_{\theta}^2c_{\alpha}^2)\mathrm{d}\varphi_3\mathrm{d}\psi\nonumber\\
& &\ \ \ \ \ \ \ \ \ \ \ \ \ \ \ \ \ \ \ \ \ \ \ \  +\Big(1+\tilde{\gamma}^2(2^5\pi^2N)f_1\Delta'_{123}\Big)\mathrm{d}\psi^2\Big\}\Bigg]
\end{eqnarray}
with 
\begin{eqnarray}
F^{(4)}_{\gamma}&= &l_p^3 \ \Bigg[\frac{3\rho u^2}{(2^5\pi^2N)} \ \mathrm{d}u\wedge\mathrm{d}t\wedge\mathrm{d}\rho\wedge\mathrm{d}\varphi_4\nonumber\\
& &\ \ \ \ \ \ +6(2^5\pi^2N) \tilde{\gamma}\sqrt{\Delta'_{123}}s_{\theta}^2s_{\alpha} \ \mathrm{d}\alpha\wedge\mathrm{d}\beta\wedge\mathrm{d}\theta\wedge\mathrm{d}\psi\nonumber\\
& &\ \ \ \ \ \ -\sum_{i\in\{\alpha,\beta,\theta\}}\frac{\tilde{\gamma}(2^5\pi^2N)\partial_i \Delta'_{123}}{(1+\tilde{\gamma}^2(2^5\pi^2N)\Delta'_{123})^2} \ \mathrm{d}x^{i}\wedge\mathrm{d}\varphi_1\wedge\mathrm{d}\varphi_2\wedge\mathrm{d}\varphi_3\Bigg]
\end{eqnarray}

The near horizon region of the undeformed membrane is 
\begin{equation}
\mathrm{ds}^2=\mathrm{ds}^2_{\mathrm{AdS}_4}(\frac{1}{2}L)+\mathrm{d}\Omega_7^2(L)\end{equation}
where
\begin{equation}
L^2=l_p^2(2^5\pi^2N)^{1/3}
\end{equation}
To see the effect of the deformation, the near horizon solution can be expressed 
in terms of the undeformed near horizon solution
\begin{eqnarray}
\mathrm{ds}^2_{\gamma}&=&(1+\tilde{\gamma}^2(2^5\pi^2N) \Delta'_{123})^{1/3}\mathrm{ds}^2_{\mathrm{AdS}_4}(\frac{1}{2}L)+\frac{1}{(1+\tilde{\gamma}^2(2^5\pi^2N) \Delta'_{123})^{2/3}}\mathrm{d}\Omega^2_{7}(L)\nonumber\\
&+&\tilde{\gamma}^2\frac{l_p^2(2^5\pi^2N)^{4/3}\Delta'_{123}}{(1+\tilde{\gamma}^2(2^5\pi^2N)\Delta'_{123})^{2/3}}\Big\{\mathrm{d}\theta^2+s_{\theta}^2(\mathrm{d}\alpha^2+s_{\alpha}^2\mathrm{d}\beta^2) + f\mathrm{d}\psi^2\Big\}
\end{eqnarray}
While the field strength for the undeformed brane in the near horizon limit is
\begin{equation}
F^{(4)}_0\sim l_p^3 \ \frac{3\rho
  u^2}{(2^5\pi^2N)}\mathrm{d}u\wedge\mathrm{d}t\wedge\mathrm{d}\rho\wedge\mathrm{d}\varphi_4 \, .
\end{equation}

This is also the solution which is obtained when one starts with the
membrane and takes the near horizon solution first and then performs a
deformation of that solution. The deformation process and decoupling
limit therefore commute. It is not clear whether this is inevitable. 
This result continues to hold true for the dipole-type deformations of the membrane. 
For completeness an example solution is included in Appendix B.

\subsection{The M5 Brane}
The near horizon limit of the M5 brane is found by taking
\begin{equation}
l_p\rightarrow 0
\end{equation}
with
\begin{equation}
u^2\equiv \frac{r}{l_p^3}=\mathrm{fixed} \, .
\end{equation}
This limit allows us to explore the near horizon region while keeping
the energies on the brane world volume fixed.
As for the membrane, finiteness of the Einstein-Hilbert term requires
$ds^2$ to be finite in units of the planck length 
squared. We also wish to preserve the deformation in the near horizon limit. 
Appearing in the deformed metric are terms like
\begin{equation}
H^{-1/3}(1+\gamma^2\Delta), \ \
H^{2/3}(1+\gamma^2\Delta)^{1/3}\mathrm{d}r^2, \ \
H^{2/3}(1+\gamma^2\Delta)^{1/3}r^2,\ \
H^{-1/3}(1+\gamma^2\Delta)^{-2/3} \, .
\end{equation}
As for the membrane, the only condition that must be satisfied to ensure the Einstein-Hilbert term is finite
in the near horizon limit is
\begin{equation}
1+\gamma^2\Delta=\mathrm{fixed} \, .
\end{equation}
Every possible deformation scenario leads to a scaling of the deformation parameter
like
\begin{equation}
\gamma\sim\tilde{\gamma} \ l_p^{-3}  \, .
\end{equation}
The non-vanishing field strength components then scale like the planck
length cubed as for the membrane and as required for a finite supergravity
action. Firstly, we present the five brane analogue of the non-commutative deformation in the decoupling limit.
Again for simplicity we work with the magnitudes and polar angles of the complexified coordinate system.
\begin{eqnarray}
\mathrm{ds}^2_{\gamma}&=&l_p^2\Bigg\{(1+\tilde{\gamma}^2(\frac{u^6}{N\pi})\mu_1^2\mu_2^2\mu_3^2)^{1/3}\Big(\frac{u^2}{(N\pi)^{1/3}}\sum_{i=1}^3\mathrm{d}\mu_i^2+4\frac{(N\pi)^{2/3}}{u^2}\mathrm{d}u^2+(N\pi)^{2/3}\mathrm{d}\Omega_4^2\Big)\nonumber\\
& &\ \ \ \ \ \ \ \ \ \ +
\frac{1}{(1+\tilde{\gamma}^2(u^6/N\pi)\mu_1^2\mu_2^2\mu_3^2)^{2/3}} \
\frac{u^2}{(N\pi)^{1/3}}\mu_i^2\mathrm{d}\phi_i^2 \Bigg\} \, .
\end{eqnarray}
All terms within the four form remain in the decoupling limit. We find that 
\begin{equation}
F_{0}^{(4)}= -3l_p^3\tilde{r}_1\tilde{r}_2(N\pi)\mathrm{d}\tilde{r}_1\wedge\mathrm{d}\theta_1\wedge\mathrm{d}\tilde{r}_2\wedge\mathrm{d}\theta_2
\end{equation}
whilst the terms originating with the three-form potential turned on
along the $T^3$ can depend on $r$ in one of two possible ways, both of
which scale in such a way that the resultant field strenght components
are finite in units of $l_p^3$ in the near horizon. That is:
\begin{eqnarray}
F_{u\phi_1\phi_2\phi_3}^{\gamma}&= & \partial_{u}H^{-1} \sim l_p^3\nonumber\\
F_{\kappa\phi_1\phi_2\phi_3}^{\gamma}&= &H^{-1}\sim l_p^3\nonumber
\end{eqnarray}
for $\kappa\in\{\mu_1, \mu_2, \mu_3\}$. At the same time, the term
generated via the deformation procedure as a result of non-vanishing
$C_{\mu\nu\lambda}$ in the IIB solution is
\begin{displaymath}
\gamma\sqrt{\Delta}\star_8F_{0}^{(4)}
\end{displaymath}
which scales in the same way as $F_{0}^{(4)}$ in the decouling limit since the terms involved scale like
\begin{displaymath}
\gamma\sqrt{H^{-1}}\sim  l_p^0
\end{displaymath}
and
\begin{displaymath}
\sqrt{|\mathrm{det}g_{(8)}|}G^{\tilde{r}_1\tilde{r}_1}G^{\theta_1\theta_1}G^{\tilde{r}_2\tilde{r}_2}G^{\theta_2\theta_2}\sim  l_p^0
\end{displaymath}
giving
\begin{equation}
\gamma\sqrt{\Delta}\star_8F_{0}^{(4)}\sim F_{0}^{(4)}
\end{equation}
which is finite in units of $l_p^3$. 
For the membrane analogue of the dipole deformations of type I using $T^3$ on $\{\phi_2, \phi_3, \theta_1\}$the decoupling limit gives
\begin{eqnarray}
\mathrm{ds}^2_{\gamma}&= & l_p^2\Bigg\{ \frac{1}{(1+\tilde{\gamma}^2u^4\tilde{r}_1^2\mu_2^2\mu_3^2)^{2/3}}\Big(\frac{u^2}{(N\pi)^{1/3}}\sum_{i=2}^3\mu_i^2\mathrm{d}\phi_i^2 + (N\pi)^{2/3}\tilde{r}_1^2\mathrm{d}\theta_1^2\Big)\nonumber\\
& & \ \ \ +(1+\tilde{\gamma}^2u^4\tilde{r}_1^2\mu_2^2\mu_3^2)^{1/3}\Bigg(\frac{u^2}{(N\pi)^{1/3}}\big(\sum_{i=1}^3\mathrm{d}\mu_i^2+\mu_1^2\mathrm{d}\phi_1^2\big)+4\frac{(N\pi)^{2/3}}{u^2}\mathrm{d}u^2\\
& & \ \ \ \ \ \ \ \ \ \ \ \  +(N\pi)^{2/3}\bigg(\tilde{r}_2^2\mathrm{d}\theta_2^2+\frac{(1-\tilde{r}_2^2)}{(1-\tilde{r}_1^2-\tilde{r}_2^2)}\mathrm{d}\tilde{r}_1^2+\frac{(1-\tilde{r}_1)}{(1-\tilde{r}_1^2-\tilde{r}_2^2)}\mathrm{d}\tilde{r}_2^2+\frac{2\tilde{r}_1\tilde{r}_2}{(1-\tilde{r}_1^2-\tilde{r}_2^2)}\bigg) \Bigg) \Bigg\}\nonumber\\
F_{\gamma}^{(4)}&= & l_p^3\Big\{-3\tilde{r}_1\tilde{r}_2(N\pi)\mathrm{d}\tilde{r}_1\wedge\mathrm{d}\theta_1\wedge\mathrm{d}\tilde{r}_2\wedge\mathrm{d}\theta_2\nonumber\\
& &\ \ \ \ \ -\frac{\tilde{\gamma}}{(1+\tilde{\gamma}^2\tilde{r}_1^2u^4\mu_2^2\mu_3^2)^2}\Big(4u^3\tilde{r}_1^2\mu_2^2\mu_3^2\mathrm{d}u\wedge\mathrm{d}\phi_2\wedge\mathrm{d}\phi_3\wedge\mathrm{d}\theta_1\nonumber\\
& & \ \ \ \ \ \ \ \ \ \ \ \ \ \ \ \ \ \ \ \ \ \ \ \ \ \ \ \ \ \ +u^4\sum_{\kappa}\partial_{\kappa}(\tilde{r}_1^2\mu_2^2\mu_3^2)\mathrm{d}x^{\kappa}\wedge\mathrm{d}\phi_2\wedge\mathrm{d}\phi_3\wedge\mathrm{d}\theta_1\Big)\Big\}
\end{eqnarray}
with $\kappa\in\{\tilde{r}_1,\mu_2,\mu_3\}$.
For the five brane dipole deformations of type II we have for all choices of reduction and T-duality
\begin{eqnarray}
F_{\gamma}^{(4)}&=&F_{0}^{(4)}-\gamma\sum_{\kappa}\frac{\partial_{\kappa}\Delta}{(1+\gamma^2\Delta)^2}\mathrm{d}x^{\kappa}\wedge\mathrm{d}\varphi_1\wedge\mathrm{d}\theta_1\wedge\mathrm{d}\theta_2\nonumber\\
&\sim & -l_p^3\Bigg\{3(N\pi)\tilde{r}_1^2\tilde{r}_2^2\mathrm{d}\tilde{r}_1\wedge\mathrm{d}\theta_1\wedge\mathrm{d}\tilde{r}_2\wedge\mathrm{d}\theta_2\nonumber\\
&+&\frac{\tilde{\gamma}}{(1+\tilde{\gamma}^2u^2(N\pi)\Delta')^2}\Big(2(N\pi)u\tilde{r}_1^2\tilde{r}_2^2\rho^2s_{\alpha}^2\mathrm{d}u\wedge\mathrm{d}\varphi_1\wedge\mathrm{d}\theta_1\wedge\mathrm{d}\theta_2\nonumber\\
&+&u^2(N\pi)\sum_{m}\partial_{\kappa}\Delta'\mathrm{d}x^{\kappa}\wedge\mathrm{d}\varphi_1\wedge\mathrm{d}\theta_1\wedge\mathrm{d}\theta_2\Big)\Bigg\}
\end{eqnarray}
for $m\in\{\rho,\alpha,\tilde{r}_1,\tilde{r}_2\}$. 
For the metric we have already seen 
\begin{eqnarray}
\mathrm{ds}^2_{\gamma}&=&\frac{1}{(1+\gamma^2\Delta)^{2/3}}\Big\{\overbrace{\mathrm{ds}_0^2}^{(1)}+\overbrace{\Delta^{-1/6}(\gamma^2\Delta)g_{\mu\nu}\mathrm{d}x^{\mu}\mathrm{d}x^{\nu}}^{(2)}\nonumber\\
& &\ \ \ \ \ \ \ \ \ \ \ \ \ \ \ \ \ \ \ +\underbrace{\Delta^{1/3}M_{\varphi_1\varphi_1}(\delta^{\varphi_1})^2}_{(3)}+\underbrace{2(-1)^P\Delta^{1/3}M_{\varphi_1\varphi_1}\mathcal{D}\varphi_1\delta^{\varphi_1}}_{(4)}\Big\} \ .
\end{eqnarray}
Given that the scaling of $\gamma$ was chosen in such a way as to preserve the deformation in the near horizon limit
\begin{displaymath}
\gamma^2\Delta \sim l_p^0 \, .
\end{displaymath}
The form in which this deformed metric has been presented allows us to quickly identify the troublesome terms by separating them out from the well behaved, familiar, terms of the undeformed metric. Namely for term $(1)$ we have
\begin{displaymath}
\mathrm{ds}^2_0\sim l_p^2
\end{displaymath}
as required. From term $(2)$, after identifying the scale independent $(\gamma^2\Delta)$ we see the familiar terms  
\begin{displaymath}
\Delta^{-1/6}g_{\mu\nu}\mathrm{d}x^{\mu}\mathrm{d}x^{\nu}\sim l_p^2
\end{displaymath}
Thus $(2)$ scales in such a way that the action remains finite as the decoupling limit is taken. After we have identified the undeformed metric term
\begin{displaymath}
\Delta^{1/3}M_{\varphi_1\varphi_1}\sim l_p^2
\end{displaymath} 
we see that terms $(3)$ and $(4)$ will scale like $(\delta^{\varphi_1})^2$ and $(\delta^{\varphi_1})$ 
respectively, whilst, as the eleven dimensional planck length is taken to zero $\delta^{\varphi^1}$ scales like
\begin{equation}
-(\frac{3N\pi}{4}) \tilde{\gamma}  \ \tilde{r}_1\tilde{r}_2\big(\tilde{r}_2\mathrm{d}\tilde{r}_1-\tilde{r}_1\mathrm{d}\tilde{r}_2\big)
\end{equation}
which produces the correct scaling of the metric in the near horizon limit.

\section{Probe branes}

When searching for supersymmetric configurations, a good place to start is
with the zeros of the potential from the probe brane sigma model. This is also
related to calculating the action of Wilson surfaces in the dual
theory. The probe membrane action is given by:
\begin{equation}
S_{\sigma}=\int\mathrm{d}^3\xi\Big\{\sqrt{\vert\mathrm{det}\frac{\partial
    x^m}{\partial x^i}\frac{\partial x^n}{\partial
    x^j}g_{mn}\vert}-\frac{1}{6}\varepsilon^{ijk}\frac{\partial
  x^m}{\partial x^i}\frac{\partial x^n}{\partial{x^j}}\frac{\partial
  x^p}{\partial x^k}C_{mnp}\Big\} \, .
\end{equation}

We first orient the probe brane to be parallel to the original brane stack.
Then, after choosing static or Monge gauge we look for brane
configurations that minimise the potential energy of the brane. 
This gives the following possiblities for each type of deformation. 
The potential vanishes for
\begin{eqnarray}
T^3  \ \ \mathrm{on}  \ \ \{\varphi_1,\varphi_2,\varphi_3\}&\rightarrow&\alpha=0, \ \ \mathrm{or} \ \ \ \theta=0\nonumber\\
T^3  \ \ \mathrm{on}  \ \ \{\varphi_4,\varphi_1,\varphi_2\}&\rightarrow&\alpha=0, \ \ \mathrm{or} \ \ \ \theta=0\nonumber\\
T^3 \ \ \mathrm{on}   \ \ \{\varphi_4,\varphi_1,\varphi_3\}&\rightarrow&\alpha=0, \ \ \mathrm{or} \ \ \ \theta=0\nonumber\\
T^3 \ \ \mathrm{on}   \ \ \{\varphi_4,\varphi_2,\varphi_3\}&\rightarrow&\theta=0 \ \ \mathrm{only}\nonumber
\end{eqnarray}
The supersymmetric locus for the membrane deformed on
$\{\varphi_4,\varphi_2,\varphi_3\}$ is zero dimensional unlike each of
the other three cases. The particular $T^3$ embedding chosen has
physical consequences. 

For the five-brane, when the deformation has
$C^{(3)}$ vanishing on the world volume, we can take the simple $\sigma$-model
\begin{equation}
S_{\sigma}=\int\mathrm{d}^6x\Big\{\sqrt{|\mathrm{det}\frac{\partial x^m}{\partial x^i}\frac{\partial x^n}{\partial x^j}g_{mn}|}-f^{*}(C^{(6)})\Big\}
\end{equation}
where $f^*$ is the pull back operation. Again, loooking for the vanishing
of the sigma model potential after orienting the probe along the original stack reveals that the condition for minimising the potential is equivalent to the vanishing of the functions $\Delta_{\varphi_i \varphi_j \varphi_k}$ given below:
\begin{eqnarray}
\Delta_{\varphi_1\varphi_2\theta_1}&=&r^2\tilde{r}_1^2\rho^4s_{\alpha}^2\big(c_{\alpha}^2+\frac{1}{4}s_{\alpha}^2s_{2\beta}^2\big)\nonumber\\
\Delta_{\varphi_1\varphi_2\theta_2}&=&r^2\tilde{r}_2^2\rho^4s_{\alpha}^2\big(c_{\alpha}^2+\frac{1}{4}s_{\alpha}^2s_{2\beta}^2\big)\nonumber\\
\Delta_{\varphi_1\theta_1\theta_2}&=&H\rho^2r^4\tilde{r}_1^2\tilde{r}_2^2s_{\alpha}^2\nonumber\\
\Delta_{\varphi_2\theta_1\theta_2}&=&H\rho^2r^4\tilde{r}_1^2\tilde{r}_2^2\big(c_{\alpha}^2+s_{\alpha}^2c_{\beta}^2)
\, . \nonumber
\end{eqnarray}
As $\alpha,\beta\in(0,\frac{\pi}{2})$, deformations on the first three $T^3$'s listed leave only the point $\alpha=0$ in the supersymmetric locus. However, for $T^3$ on $\{\varphi_2,\theta_1,\theta_2\}$ there exist configurations, which are supersymmetric for $\alpha=0$ with any value of $\beta$ or $\beta=0$ with any value of $\alpha$.

\section{Black Branes, $\gamma$-Deformations and Entropy}

We now consider non-extremal versions of thes deformed solutions. That
is we will thermalise the deformed branes and analyse their thermodynamics.
The Bekenstein-Hawking Entropy of the non-extremal 
deformed solutions is given by \cite{duff}:
\begin{equation}
S_{\mathcal{BH}}=\frac{1}{4}\int\sqrt{\mathrm{det}g'_{ij}}\mathrm{d}^9x
\end{equation}        
where the prime indicates that it is the reduced determinant. That is
the determinent taken over 
the nine-dimensional block of the metric where temporal and 
radial components are neglected.

We will examine the effects of the deformation on the black brane
entropy for the various types of deformation. 

Following, \cite{duff}, we thermalise the deformed branes with the
introduction of the following factor and its
reciprocal into the temporal and radial components of the metric as follows:
\begin{eqnarray}
G_{tt}&\rightarrow& (1-{{r_0 \over r^c}})         G_{tt}\nonumber\\
G_{rr}&\rightarrow& (1-{{r_0 \over r^c}})^{-1}         G_{rr}\nonumber
\end{eqnarray} 
where c is the power of r appearing in the brane harmonic function
(this is obviously dependent on the dimension of the brane. (One must
also change the definition on the brane charge \cite{duff}).
As with the study of Probe branes and supersymmetry, 
the particular embedding of the $T^3$ is important in this analysis.

For a $T^3$ trivially fibred 
over an 8 dimensional manifold, the metric may be placed in block diagonal form 
(with one $3\times 3$ block 
describing the $T^3$ and the $8\times 8$ block describing the perpendicular
space). The $T^3$ volume is 
warped homogenously by the deformation as is the eight dimensional space. 
This occurrs in such a way as to cancel the effect of the warping on the
determinant of the reduced metric used in the Bekenstein-Hawking formula.
The deformed metric, $G^{(\gamma)}_{MN}$ may
be written in terms of the nondeformed metric $G^{(0)}_{MN}$ as follows:
\begin{equation}
G_{MN}^{(\gamma)}=\begin{pmatrix} (1+\gamma^2\Delta)^{-2/3}G_{ab}^{(0)} & 0 \\ 0 & (1+\gamma^2\Delta)^{1/3}G_{\mu\nu}^{(0)}\end{pmatrix} 
\end{equation}
where $a,b=1,2,3,\ \mu,\nu=4\cdots 10$. This being the only change, the 
multiplicative factors can be pulled outside to produce an overall 
multiplicative factor in the determinant. Specifically, calculating the reduced
determinant for the Bekenstein-Hawking Entropy gives 
\begin{equation}
\mathrm{det'}G_{MN}^{(\gamma)}=[(1+\gamma^2\Delta)^{-2/3}]^{3}\times[(1+\gamma^2\Delta)^{1/3}]^{11-3-2}\times\mathrm{det'}G_{MN}^{(0)}
\end{equation}
And so for the M-theory analogues (where they apply) of the Non-Commutative, 
Dipole and $\gamma$-deformations of the Membrane 
\begin{equation}
\mathrm{det'}G_{MN}^{(\gamma)}=\mathrm{det'}G_{MN}^{(0)}
\end{equation}
and therefore
\begin{equation}
S_{\mathcal{BH}}^{(\gamma)}=S_{\mathcal{BH}}^{(0)} \, .
\end{equation}

Thus in fact there is no change in entropy. The deformed brane has the
same entropy as the undeformed 
brane.\footnote{This is provided that the temporal direction is not included in the
deformation procedure (which not always the case for our the five-brane).} 

This invariance of the entropy under the deformation procedure for trivial embeddings is a
consequence of the properties of this specific deformation in eleven
dimensions. It is not the case for the analagous deformation of a ten dimensional 
Supergravity theory using a $U(1)\times U(1)$ background isometry. 

This is quite unusual, since the deformation will introduce an
interaction in the dual description and so one would imagine that 
the entropy of the theory would change. It is possible that this is a consequence of the
large N limit, since the supergravity description is only good at
large N. Perhaps it is possible that the deformations of the theory
are somehow subleading at large N but this seems unlikey. The original
Lunin Maldaence deformation corresponding to the beta deformed theory
certianly had large N implications. Also, the solution is deformed,
only the reduced determinant (which is relevent for the entropy) is invariant.

Of the deformations of these eleven dimensional 
backgrounds with trivial $T^3$ fibration, there is however one special 
example not sharing this cancellation. The ``Non-Commutative'' 
deformation of the five-brane world-volume produces a metric with a  
reduced determinant that retains a remnant of the deformation procedure. 

The simplicity of the general cases of deformation of trivial
fibrations 
and the observed invariance of the 
Bekenstein-Hawking entropy is in marked contrast to the effect
of the deformation involving the choice of a non-trivial $T^3$ embedding. 
Geometrically along with the $T^3$ expansion/contraction, 
the eight dimensional base in these cases 
experiences a form of shearing. The deformation takes effect
in a highly inhomogeneous manner. The result of this is the appearence of 
additional additive terms within the determinant of the metric. 
These are the source of changes in the Bekenstein-Hawking Entropy.     
As an example we consider the deformation of the membrane on $\{\varphi_1,\varphi_2,\varphi_3\}$ where
\begin{equation}
\sqrt{|\mathrm{det}'G_{MN}^{(\gamma)}|}=\frac{1}{(1+\gamma^2\Delta_{123})^{1/2}}\sqrt{|\mathrm{det}'G_{MN}^{(0)}+\gamma^2\rho^2r^8s_{\theta}^4s_{\alpha}^2\Delta_{123}^2f_1|}
\end{equation}
A numerical evaluation was attempted but as yet, an explicit solution has not presented itself.
For the deformations involving $T^3$'s with one-cycles both on and off the membrane world volume, the changes in the reduced determinant are far more compilcated. for each of these cases we have in general
\begin{equation}
\sqrt{|\mathrm{det}'G_{MN}^{(\gamma)}|}=\frac{1}{(1+\gamma^2\Delta)}\sqrt{|\mathrm{det}'G_{MN}^{(0)}+\mathcal{O}(\gamma^2)+\mathcal{O}(\gamma^4)|}
\end{equation}
Along with the larger power of the deformation pre-factor in the denominator, there are far more terms at $\mathcal{O}(\gamma^2)$ inside the square root. The more complicated the embedding, the more no-trivial the functions making an appearance here. The $\mathcal{O}(\gamma^4)$ are a new feature in the reduced determinant for the ``dipole" type deformations of the membrane. For deformation on 
$\{\varphi_4,\varphi_1,\varphi_2\}$
\begin{eqnarray}
\mathrm{det}'G_{MN}^{(\gamma)}&=&\frac{1}{(1+\gamma^2\Delta_{412})}\Bigg\{\mathrm{det}'G_{MN}^{(0)}\nonumber\\
&+&\gamma^2\Big[(\rho^2r^6s_{\theta}^4s_{\alpha}^2\Delta_{412}^2)\Big(g_1G_{\varphi_3\varphi_3}^{(0)}+g_2G_{\psi\psi}^{(0)}-2g_3G_{\varphi_3\psi}^{(0)}\Big)\Big]\nonumber\\
&+&\gamma^4\Big[H^{1/3}(r^2\Delta_{412})^2r^6\rho^2s_{\theta}^4s_{\alpha}^2\Big(g_1g_2-g_3^2\Big)\Delta_{412}\Big]\Bigg\}
\end{eqnarray}

\section{Conclusions}

This paper has explored the result of applying the M-theory deformation process
to branes. In particular, we have seen how to generate families of
solutions beginning with the usual membrane and five-brane solutions. 
The deformed solutions have a well defined decoupling limit 
where the deformation is scaled so as to survive the decoupling limit.
Importantly, the deformation process is shown to commute with taking
decoupling limit.

The properties of these solutions were investigated. Perhaps the most
intruiging result is that the entropy of a class of deformed branes is the
same as that of the undeformed branes despite the deformed theory
having different amounts of supersymmetry and new interactions. 

Let us consider the membrane. From the dual theory perspective this is
a statment that there is for N=8, 2+1 dimensional Yang-Mills theory in
the large N limit, at strong
coupling there is a deformation that breaks supersymmetry and yet
preserves the entropy of the theory.

There are also some deeper questions about M-theory geometry. We know
from other work how the so called ``open string'' metric is an
invariant under a similar deformation process \cite{def1}. Later it was show that
one could form the equivalent invariant metric for M-theory
\cite{def2}. It would be interesting to see if one could form a similar
invariant combination of metric and C-field that captures the class of
the solution and is invariant under the deformation process. This is
similar in spirit to the generalised geometry prgram where
T-dualities are just diffeomorphisms of a more general geometry of
bigger space. Perhaps the entropy invariance mentioned above is
a sign of such a deeper symmetry.

One future possibility would be to look at $G_2$ manifolds which have
interesting phenomenological possibilities and study the deformations
of these. This may allow the introduction of a new scale (coming from
the deformation) that may allow additional decoupling.

\section{Acknowledgements}

DSB is supported by EPSRC grant GR/R75373/02 and would like to thank DAMTP and Clare Hall college
Cambridge for continued support. This work was in part supported by the EC Marie Curie
Research Training Network, MRTN-CT-2004-512194. LT is supported by a PPARC grant and would like to thank Dario Duo for many helpful conversations.

\begin{appendix}
\section{``Dipole'' deformations of the membrane}
\subsection{$T^3$ on $\{\varphi_4,\varphi_1,\varphi_2\}$}
The deformed M2 using $T^3$ given by $\{\varphi_4,\varphi_1,\varphi_2\}$ is
\footnote{\begin{equation}csc_{\theta}=cosec[\theta], sec_{\theta}=sec[\theta]\end{equation}}

\begin{eqnarray}
\mathrm{ds}^2_{\gamma} &=& (1+\gamma^2\Delta_{412})^{1/3}\Bigg\{H^{-2/3}\Big(-\mathrm{d}t^2+\mathrm{d}\rho^2\Big)+H^{1/3}\Big(\mathrm{d}r^2\nonumber\\
& & \ \ \ \ \ \ \ \ \ \ \ \ \ \ \ \ \ \ \ \ \ \ \ \ \ \ \ \ \ \ \ \ \ \ \ \ \ \ \ \ \  +r^2(\mathrm{d}\theta^2+s_{\theta}^2(\mathrm{d}\alpha^2+s_{\alpha}^2\mathrm{d}\beta^2))\Big)\Bigg\}\nonumber\\
&+&\frac{1}{(1+\gamma^2\Delta_{412})^{2/3}}\Bigg\{H^{-2/3}\rho^2\mathrm{d}\varphi_4^2+H^{1/3}r^2\Big(\nonumber\\
& &\ \ \ \ \ \ \ \ \ \ \ \ \ \ \ \ \ \ \ \ \ \ \ \ \ \ \ \ \ \ \ s_{\theta}^2s_{\alpha}^2\mathrm{d}\varphi_1^2+s_{\theta}^2(s_{\alpha}^2c_{\beta}^2+c_{\alpha}^2)\mathrm{d}\varphi_2^2\nonumber\\
& & \ \ \ \ \ \ \ \ \ \ \ \ \ \ \ \ \ \ \ \ \ \ \ \ \ \ \ \ \ \ -2_{\beta}^2s_{\theta}^2s_{\alpha}^2\mathrm{d}\varphi_1\mathrm{d}\varphi_2 +2_{\alpha}^2s_{\theta}^2(s_{\beta}^2-c_{\beta}^2)\mathrm{d}\varphi_1\mathrm{d}\psi\nonumber\\
& &\ \ \ \ \ \ \ \ \ \ \ \ \ \ \ \ \ \ \ \ \ \ \ \ \ \ \ \ \ \ \ +2_{\theta}^2(s_{\alpha}^2c_{\beta}^2-c_{\alpha}^2)\mathrm{d}\varphi_2\mathrm{d}\psi + 2_{\theta}^2c_{\alpha}^2\mathrm{d}\varphi_{2}\mathrm{d}\varphi_{3}\Big)\Bigg\}\nonumber\\
&+&\frac{H^{1/3}r^2}{(1+\gamma^2\Delta_{412})^{2/3}}\Bigg\{\Big(1+\gamma^2\Delta_{412} \ g_1(\alpha,\beta,\theta)\Big)\mathrm{d}\psi^2\nonumber\\
& &\ \ \ \ \ \ \ \ \ \ \ \ \ \ \ \ \ \ \ \ \ \ +\Big((c_{\theta}^2+s_{\theta}^2c_{\alpha}^2)+\gamma^2\Delta_{412} \ g_2(\alpha,\beta,\theta)\Big)\mathrm{d}\varphi_3^2\nonumber\\
& &\ \ \ \ \ \ \ \ \ \ \ \ \ \ \ \ \ \ \ \ \ \ +\Big((c_{\theta}^2-s_{\theta}^2c_{\alpha}^2)+\frac{1}{2}\gamma^2\Delta_{412} \ g_3(\alpha,\beta,\theta)\Big)\mathrm{d}\psi\mathrm{d}\varphi_{3}\Bigg\}\\
F^{(4)}_{\gamma}&=& F^{(4)}_0 -\partial_{\kappa}\Big[\frac{\gamma\Delta_{412}}{(1+\gamma^2\Delta_{412})}\Big]\mathrm{d}x^{\kappa}\wedge\mathrm{d}\varphi_4^2\wedge\mathrm{d}\varphi_1^2\wedge\mathrm{d}\varphi_2^2
\end{eqnarray}
where $\kappa\in\{\rho,r,\alpha,\beta,\theta\}$ and
\begin{equation}
\Delta_{412}=\rho^2 r^4\Big(s_{\theta}^4s_{\alpha}^2\big(s_{\alpha}^2c_{\beta}^2+c_{\alpha}^2\big)-s_{\theta}^4c_{\beta}^4s_{\alpha}^4\Big)
\end{equation}
with
\begin{eqnarray}
g_1(\alpha,\beta,\theta)&=&\frac{16c_{\alpha}^2(1-(c_{\alpha}^2+c_{4\beta}s_{\alpha}^2)s_{\theta}^2}{(9+7c_{2\alpha}-2c_{4\beta}s_{\alpha}^2)}\nonumber\\  
g_2(\alpha,\beta,\theta)&=&c_{\theta}^2+\frac{s_{\theta}^2}{4csc_{\alpha}^2 \ csc_{2\beta}^2+sec_{\alpha}^2}\nonumber\\
g_3(\alpha,\beta,\theta)&=&2c_{\theta}^2-\frac{c_{\alpha}^2(3+c_{2\beta})s_{\alpha}^2s_{\beta}^2s_{\theta}^2}{(c_{\alpha}^2+c_{\beta}^2s_{\alpha}^2s_{\beta}^2)}\nonumber
\end{eqnarray}

\subsection{$T^3$ on $\{\varphi_4,\varphi_1,\varphi_3\}$}
The deformed M2 using $T^3$ with $\{\varphi_4,\varphi_1,\varphi_3\}$ is
\begin{eqnarray}
\mathrm{ds}^2 &=& (1+\gamma^2\Delta_{413})^{1/3}\Bigg\{H^{-2/3}\Big(-\mathrm{d}t^2+\mathrm{d}\rho^2\Big)+H^{1/3}\Big(\mathrm{d}r^2\nonumber\\
& &\ \ \ \ \ \ \ \ \ \ \ \ \ \ \ \ \ \ \ \ \ \ +r^2(\mathrm{d}\theta^2+s_{\theta}^2(\mathrm{d}\alpha^2+s_{\alpha}^2\mathrm{d}\beta^2))\Big)\Bigg\}\nonumber\\
&+&\frac{1}{(1+\gamma^2\Delta_{413})^{2/3}}\Bigg\{H^{-2/3}\rho^2\mathrm{d}\varphi_4^2+H^{1/3}r^2\Big(s_{\theta}^2s_{\alpha}^2\mathrm{d}\varphi_1^2\nonumber\\
& &\ \ \ \ \ \ \ \ \ \ \ \ \ \ \ \ \ \ \ \ \ \ +(c_{\theta}^2+s_{\theta}^2c_{\alpha}^2)\mathrm{d}\varphi_3^2+s_{\alpha}^2s_{\theta}^2(s_{\beta}^2-c_{\beta}^2)\mathrm{d}\varphi_1\mathrm{d}\psi\nonumber\\
& &\ \ \ \ \ \ \ \ \ \ \ \ \ \ \ \ \ \ \ \ \ \ +(c_{\theta}^2-s_{\theta}^2c_{\alpha}^2)\mathrm{d}\varphi_3\mathrm{d}\psi -c_{\beta}^2s_{\theta}^2s_{\alpha}^2\mathrm{d}\varphi_1\mathrm{d}\varphi_2\nonumber\\
& &\ \ \ \ \ \ \ \ \ \ \ \ \ \ \ \ \ \ \ \ \ \ +s_{\theta}^2c_{\alpha}^2\mathrm{d}\varphi_2\mathrm{d}\varphi_3\Big)\Bigg\}\nonumber\\
&+&\frac{H^{1/3}r^2}{(1+\gamma^2\Delta_{413})^{2/3}}\Bigg\{\Big(s_{\theta}^2(s_{\alpha}^2c_{\beta}^2+c_{\alpha}^2)+\gamma^2\Delta_{413} \  h_1(\alpha,\beta,\theta)\Big)\mathrm{d}\varphi_2^2\nonumber\\
& &\ \ \ \ \ \ \ \ \ \ \ \ \ \ \ \ \ \ \ \ \ +\Big(1+\gamma^2\Delta_{413} \ h_2(\alpha,\beta,\theta)\Big)\mathrm{d}\psi^2\nonumber\\
& &\ \ \ \ \ \ \ \ \ \ \ \ \ \ \ \ \ \ \ \ \ +\Big(s_{\theta}^2(s_{\alpha}^2c_{\beta}^2-c_{\alpha}^2)+\gamma^2\Delta_{413} \ h_3(\alpha,\beta,\theta)\Big)\mathrm{d}\varphi_2\mathrm{d}\psi\Bigg\}\\
& &\nonumber\\
F^{(4)}_{\gamma}&=&F^{(4)}_{0}-\partial_{\kappa} \ \Big[\frac{\gamma\Delta_{412}}{(1+\gamma^2\Delta_{413})}\Big] \ \mathrm{d}x^{\kappa}\wedge\mathrm{d}\varphi_4\wedge\mathrm{d}\varphi_1\wedge\mathrm{d}\varphi_3
\end{eqnarray}
for $\kappa\in\{\rho, r, \alpha, \beta, \theta\}$ with
\begin{equation}
\Delta_{413}=\rho^2 r^4 \Big(s_{\alpha}^2s_{\theta}^2(c_{\theta}^2+c_{\alpha}^2s_{\theta}^2)\Big)
\end{equation}
and
\begin{eqnarray}
h_1(\alpha,\beta,\theta)&=&c_{\theta}^2+c_{\beta}^2s_{\alpha}^2s_{\beta}^2s_{\theta}^2-\frac{c_{\theta}^4}{(c_{\theta}^2+c_{\alpha}^2s_{\theta}^2)}\nonumber\\
h_2(\alpha,\beta,\theta)&=&1+3c_{\theta}^2-(c_{\alpha}^2+c_{2\beta}^2s_{\alpha}^2)s_{\theta}^2-\frac{4c_{\theta}^4}{(c_{\theta}^2+c_{\alpha}^2s_{\theta}^2)}\nonumber\\
h_3(\alpha,\beta,\theta)&=&c_{\beta}^2s_{\alpha}^2s_{\beta}^2s_{\theta}^2-\frac{1}{(csc_{\theta}^2sec_{\alpha}^2+sec_{\theta}^2)}\nonumber
\end{eqnarray}

\subsection{$T^3$ on $\{\varphi_4,\varphi_2,\varphi_3\}$}
The deformed M2 using $T^3$ with $\{\varphi_4,\varphi_2, \varphi_3\}$ is
\begin{eqnarray}
\mathrm{ds}^2 &=& (1+\gamma^2\Delta_{423})^{1/3}\Bigg\{H^{-2/3}\Big(-\mathrm{d}t^2+\mathrm{d}\rho^2\Big)+H^{1/3}\Big(\mathrm{d}r^2\nonumber\\
& &\ \ \ \ \ \ \ \ \ \ \ \ \ \ \ \ \ \ \ \ \ \ \ \ \ \ \ \ \ \ \ +r^2(\mathrm{d}\theta^2+s_{\theta}^2(\mathrm{d}\alpha^2+s_{\alpha}^2\mathrm{d}\beta^2))\Big)\Bigg\}\nonumber\\
&+&\frac{1}{(1+\gamma^2\Delta_{423})^{2/3}}\Bigg\{H^{-2/3}\rho^2\mathrm{d}\varphi_4^2+H^{1/3}r^2\Big(\nonumber\\
& &\ \ \ \ \ \ \ \ \ \ \ \ \ \ \ \ \ \ \ \ \ +(c_{\theta}^2+s_{\theta}^2c_{\alpha}^2)\mathrm{d}\varphi_3^2+s_{\theta}^2(s_{\alpha}^2c_{\beta}^2+c_{\alpha}^2)\mathrm{d}\varphi_2^2\nonumber\\
& &\ \ \ \ \ \ \ \ \ \ \ \ \ \ \ \ \ \ \ \ \ +s_{\theta}^2c_{\alpha}^2\mathrm{d}\varphi_2\mathrm{d}\varphi_3+s_{\theta}^2(s_{\alpha}^2c_{\beta}^2-c_{\alpha}^2)\mathrm{d}\varphi_2\mathrm{d}\psi\nonumber\\
& &\ \ \ \ \ \ \ \ \ \ \ \ \ \ \ \ \ \ \ \ \ +(c_{\theta}^2-s_{\theta}^2c_{\alpha}^2)\mathrm{d}\varphi_3\mathrm{d}\psi\Big)\Bigg\}\nonumber\\
&+&\frac{H^{1/3}r^2}{(1+\gamma^2\Delta_{423})^{2/3}}\Bigg\{\Big(1+\gamma^2\Delta_{423}h_1(\alpha,\beta,\theta)\Big)\mathrm{d}\psi^2\nonumber\\
& &\ \ \ \ \ \ \ \ \ + \Big(s_{\theta}^2s_{\alpha}^2+\gamma^2\Delta_{423}h_2(\alpha,\beta,\theta)\Big)\mathrm{d}\varphi_1^2\nonumber\\
& &\ \ \ \ \ \ \ \ \ + \Big(s_{\alpha}^2s_{\theta}^2(s_{\beta}^2-c_{\beta}^2)+\gamma^2\Delta_{423} \ h_3(\alpha,\beta,\theta)\Big)\mathrm{d}\varphi_1\mathrm{d}\psi\Bigg\}\\
& &\nonumber\\
F^{(4)}_{\gamma}&=&F^{(4)}_0 - \gamma\partial_{\kappa} \ \Big[\frac{\Delta_{423}}{(1+\gamma^2\Delta_{423})}\Big] \ \mathrm{d}x^{\kappa}\wedge\mathrm{d}\varphi_4\wedge\mathrm{d}\varphi_3\wedge\mathrm{d}\varphi_2 
\end{eqnarray}
where
\begin{equation}
\Delta_{423}=\rho^2r^4(s_{\theta}^2(s_{\alpha}^2c_{\beta}^2+c_{\alpha}^2)(c_{\theta}^2+s_{\theta}^2c_{\alpha}^2)-s_{\theta}^4c_{\alpha}^4)
\end{equation}
and
\begin{eqnarray}
h_1(\alpha,\beta,\theta)&=&\frac{s_{\theta}^2c_{\beta}^2c_{\theta}^2s_{\alpha}^2(1-c_{\beta}^2s_{\alpha}^2)}{(c_{\beta}^2c_{\theta}^2s_{\alpha}^2+c_{\alpha}^2(c_{\theta}^2+c_{\beta}^2s_{\theta}^2s_{\alpha}^2))}\nonumber\\
&+&\frac{s_{\theta}^2\Big(c_{\alpha}^2(c_{\theta}^2+c_{\beta}^2s_{\alpha}^2(4+3c_{2\theta}-c_{\beta}^2s_{\alpha}^2s_{\theta}^2))-c_{\alpha}^4(c_{\theta}^2+c_{\beta}^2s_{\alpha}^2s_{\theta}^2)\Big)}{(c_{\beta}^2c_{\theta}^2s_{\alpha}^2+c_{\alpha}^2(c_{\theta}^2+c_{\beta}^2s_{\theta}^2s_{\alpha}^2))}\nonumber\\
& &\nonumber\\
h_2(\alpha,\beta,\theta)&=&\frac{s_{\theta}^2c_{\theta}^2\Big(-c_{\alpha}^6-c_{\alpha}^4(-4+3c_{\beta}^2)s_{\alpha}^2-c_{\alpha}^2c_{\beta}^2(-4+3c_{\beta}^2)s_{\alpha}^4+c_{\beta}^4s_{\alpha}^6s_{\beta}^2\Big)}{((2c_{\alpha}^2+c_{\beta}^2s_{\alpha}^2)^2(c_{\theta}^2+c_{\alpha}^2s_{\theta}^2))}\nonumber\\
&+&\frac{s_{\theta}^2c_{\theta}^2\Big(c_{\alpha}^2s_{\alpha}^2(-c_{\alpha}^4(-4+c_{\beta}^2)-c_{\alpha}^2c_{\beta}^2(-3+c_{2\beta})s_{\alpha}^2+c_{\beta}^4s_{\alpha}^4s_{\beta}^2)\Big)}{((2c_{\alpha}^2+c_{\beta}^2s_{\alpha}^2)^2(c_{\theta}^2+c_{\alpha}^2s_{\theta}^2))}\nonumber\\
& &\nonumber\\
h_3(\alpha,\beta,\theta)&=&\frac{s_{\theta}^2c_{\theta}^2(-4c_{\alpha}^4-c_{\alpha}^2(1+5c_{2\beta})s_{\alpha}^2+2c_{\beta}^2s_{\alpha}^4s_{\beta}^2)}{2(2c_{\alpha}^2+c_{\beta}^2s_{\alpha}^2)(c_{\theta}^2+c_{\alpha}^2s_{\theta}^2)}\nonumber\\
&+&\frac{s_{\theta}^2c_{\alpha}^2s_{\alpha}^2(c_{\alpha}^2(1-3c_{2\beta})+2c_{\beta}^2s_{\alpha}^2s_{\beta}^2)s_{\theta}^2)}{2(2c_{\alpha}^2+c_{\beta}^2s_{\alpha}^2)(c_{\theta}^2+c_{\alpha}^2s_{\theta}^2)}
\end{eqnarray}

\section{Near Horizon region for Membrane deformed on $\{\varphi_4,\varphi_1,\varphi_3\}$}
For $T^3$ given by $\{\varphi_4,\varphi_1,\varphi_3\}$ the near horizon
solution is
\begin{eqnarray}
\mathrm{ds}^2&{\sim\atop{l_p\rightarrow 0}}& l_p^2 \ \Bigg[(1+\tilde{\gamma}^2\rho^2u^2\Delta'_{413})^{1/3}\Bigg(\frac{u^2}{(2^4\pi^2N)^{2/3}}\Big(-\mathrm{d}t^2+\mathrm{d}\rho^2\Big)+(2^5\pi^2N)^{1/3}\Big(\nonumber\\
& &\ \ \ \ \ \ \ \ \ \ \ \ \ \ \ \ \ \ \ \ \ \ \ \ \ \ \ \ \ \ \ \ \ \ \ \ \  
\frac{\mathrm{d}u^2}{4u^2}+\big(\mathrm{d}\theta^2+s_{\theta}^2(\rm{d}\alpha^2+s_{\alpha}^2\rm{d}\beta^2)\big)\Big)\Bigg)\nonumber\\
&+&\frac{1}{(1+\tilde{\gamma}^2\rho^2 u^2\Delta'_{413})^{2/3}}\Bigg(\frac{u^2}{(2^5\pi^2N)^{2/3}}\rho^2\mathrm{d}\varphi_4^2+(2^5\pi^2N)^{1/3}\Big(\nonumber\\
& &\ \ \ \ \ \ \ \ \ \ \ \ \ \ \ \ \ \ \ \ \ \ \ \ \ \ \ \ \ \  s_{\theta}^2s_{\alpha}^2\mathrm{d}\varphi_1^2
+(c_{\theta}^2+s_{\theta}^2c_{\alpha}^2)\mathrm{d}\varphi_3^2\nonumber\\
& &\ \ \ \ \ \ \ \ \ \ \ \ \ \ \ \ \ \ \ \ \ \ \ \ \ \ \  +2s_{\alpha}^2s_{\theta}^2(s_{\beta}^2-c_{\beta}^2)\mathrm{d}\varphi_1\mathrm{d}\psi+2(c_{\theta}^2-s_{\theta}^2c_{\alpha}^2)\mathrm{d}\varphi_3\mathrm{d}\psi\nonumber\\      & &\ \ \ \ \ \ \ \ \ \ \ \ \ \ \ \ \ \ \ \ \ \ \ \ \ \ \  -2c_{\beta}^2s_{\theta}^2s_{\alpha}^2\mathrm{d}\varphi_1\mathrm{d}\varphi_2+2s_{\theta}^2c_{\alpha}^2\mathrm{d}\varphi_2\mathrm{d}\varphi_3\Big)\Bigg)\nonumber\\
&+&\frac{(2^5\pi^2N)^{1/3}}{(1+\tilde{\gamma}^2\rho^2u^2\Delta'_{413})^{2/3}}\Bigg(\Big(s_{\theta}^2(s_{\alpha}^2c_{\beta}^2+c_{\alpha}^2)+\tilde{\gamma}^2\rho^2u^2\Delta'_{413} \ h_1\Big)\mathrm{d}\varphi_2^2\nonumber\\
& &\ \ \ \ \ \ \ \ \ \ \ \ \ \ \ \ \ \ \ \ \ \ \ \ \ \ \ \ +\Big(1+\tilde{\gamma}^2\rho^2u^2\Delta'_{413} \ h_2\Big)\mathrm{d}\psi^2\nonumber\\
& &\ \ \ \ \ \ \ \ \ \ \ \ \ \ \ \ \ \ \ \ \ \ \ \ \ \ \ \  +2\Big(s_{\theta}^2(s_{\alpha}^2c_{\beta}^2-c_{\alpha}^2)+\tilde{\gamma}^2\rho^2u^2\Delta'_{413} \ h_3\Big)\mathrm{d}\varphi_2\mathrm{d}\psi\Bigg)\Bigg]\nonumber\\
F^{(4)}_{\gamma}&{\sim\atop{l_p\rightarrow 0}}&l_p^3 \ \Bigg[\frac{3\rho u^2}{2^5\pi^2 N}\mathrm{d}u\wedge\mathrm{d}t\wedge\mathrm{d}\rho\wedge\mathrm{d}\varphi_4-\frac{\tilde{\gamma}}{(1+\tilde{\gamma}^2\rho^2u^2\Delta'_{413})^2}\Bigg(\nonumber\\
& &\ \ \ \ \ \ \ \ \ \ \ \ \ \ \ \ \ \ \ \ \ \ \ \ \ \ \ \ \ \ \ \ \ \ \ \ \ \ 2u\rho^2\Delta'_{413} \ \mathrm{d}u\wedge\mathrm{d}\varphi_4\wedge\mathrm{d}\varphi_1\wedge\mathrm{d}\varphi_3\nonumber\\
& & \ \ \ \ \ \ \ \ \ \ \ \ \ \ \ \ \ \ \ \ \ \  +\sum_{i\in\{\rho\alpha,\theta\}}u^2\partial_{i}(\rho^2\Delta'_{413})\mathrm{d}x^i\wedge\mathrm{d}\varphi_4\wedge\mathrm{d}\varphi_1\wedge\mathrm{d}\varphi_3\Bigg
)\Bigg]
\end{eqnarray}
Unlike the deformation using a $T^3$ embedded entirely in the transverse space, the $F_{u\varphi_4\varphi_1\varphi_3}$ does not vanish in the decoupling limit.

\end{appendix}

\end{document}